\documentclass[journal,comsoc]{IEEEtran}

\usepackage{epsfig}
\graphicspath{{./fig/}}

\usepackage[T1]{fontenc}
\usepackage[utf8]{inputenc}
\usepackage{t1enc}

\usepackage{algorithm}
\usepackage{algorithmicx}
\usepackage{algpseudocode}
\usepackage{amsmath}
\newcommand{\COMMENT}[2][.4\linewidth]{%
	\leavevmode\hfill\makebox[#1][l]{\textcolor{gray}{\textit{\#~#2}}}}

\algnewcommand\algorithmicforeach{\textbf{for each}}
\algdef{S}[FOR]{ForEach}[1]{\algorithmicforeach\ #1\ \algorithmicdo}

\usepackage[]{units}
\usepackage{listings}
\usepackage{setspace}
\usepackage{listingsutf8}
\usepackage{array}
\usepackage{supertabular}
\usepackage{comment}
\usepackage{tabularx}
\usepackage{amsmath}
\usepackage{enumitem}
\usepackage{url}
\usepackage{xspace}
\usepackage{subfig}
\usepackage{listings}

\usepackage{tikz} 
\usepackage{pgfplots}
\pgfplotsset{compat=newest}
\usepgfplotslibrary{statistics}
\usetikzlibrary{patterns}
\usetikzlibrary{spy}

\usepackage{cleveref}
\crefname{section}{§}{§§}
\Crefname{section}{§}{§§}
\crefformat{section}{§#2#1#3}

\crefname{table}{Table}{Table}
\crefname{figure}{Fig.}{Fig.}
\crefname{listing}{Listing}{Listing}
\crefname{algorithm}{Alg.}{Alg.}

\usepackage{xcolor}
\definecolor{my_green}{HTML}{11aa33}
\colorlet{mygray}{black!30}
\colorlet{mygreen}{green!60!blue}
\colorlet{mymauve}{red!60!blue}

\usepackage{mathtools}
\DeclarePairedDelimiter\ceil{\lceil}{\rceil}

\lstdefinestyle{ovs_source}
{
  language=C, 
  backgroundcolor=\color{gray!10},  
  basicstyle=\ttfamily\scriptsize,
  columns=fullflexible,
  breakatwhitespace=false,      
  breaklines=true,                
  captionpos=b,                    
  commentstyle=\color{mygreen}, 
  extendedchars=true,              
  frame=single,                   
  keepspaces=true,             
  keywordstyle=\color{blue},      
  language=c++,                 
  numbers=none,                
  numbersep=5pt,                   
  numberstyle=\tiny\color{blue}, 
  rulecolor=\color{mygray},        
  belowskip=0pt,
  showspaces=false,                
  showstringspaces=false,          
  showtabs=false                    
  stepnumber=5,                  
  stringstyle=\color{mymauve},    
  tabsize=3,
}

\usepackage{soul}

\newcommand{\ovskernel}{\texttt{OVS-kernel}\xspace}
\newcommand{\ovssource}{\texttt{OVS-source}\xspace}
\newcommand{\ovsdpdk}{\texttt{OVS-DPDK}\xspace}

\newcommand{\naively}{na\"{\i}vely\xspace}

\ifCLASSOPTIONcompsoc
  \usepackage[nocompress]{cite}
\else
  \usepackage{cite}
\fi

\begin{document}
%
\title{On the Feasibility and Enhancement of the Tuple Space Explosion Attack against Open vSwitch}

\author{\IEEEauthorblockN{Levente Csikor, Vipul Ujawane, Dinil Mon Divakaran}
\thanks{\textit{Levente Csikor} is with the School of Computing, National University of Singapore (NUS), email: dcslev@nus.edu.sg. 
\textit{Vipul Ujawane} is with Indian Institute of Technology, Kharagpur, email: vipul999ujawane@iitkgp.ac.in. He was an intern at NUS during this work. 
\textit{Dinil Mon Divakaran} is with Trustwave (a Singtel company), email: dinil.divakaran@trustwave.com.}

}

\maketitle

\begin{abstract}
Being a crucial part of networked systems, packet classification has to be 
highly efficient; however,  software switches in cloud environments
still face performance challenges. 
The recently proposed Tuple Space Explosion (TSE) attack exploits an algorithmic deficiency in Open vSwitch (OVS).
In TSE, legitimate low-rate attack traffic makes the cardinal linear search algorithm in the Tuple Space Search (TSS) algorithm  to spend an unaffordable time for classifying each packet resulting in a denial-of-service (DoS) for the rest of the users.

In this paper, we investigate the feasibility of TSE from multiple perspectives.
Besides showing that TSE is still efficient in the newer version of OVS, we show that when the kernel datapath is compiled from a different source, it can degrade its performance to ~1\% of its baseline with less than 1 Mbps attack rate.
Finally, we show that TSE is much less effective against OVS-DPDK with userspace datapath due to the enhanced ranking process in its TSS implementation.
Therefore, we propose TSE 2.0 to defeat the ranking process and achieve a complete DoS against OVS-DPDK.
Furthermore, we present TSE 2.1, which achieves the same goal against OVS-DPDK running on multiple cores without significantly increasing the attack rate.

\end{abstract}

\IEEEpeerreviewmaketitle

\section{Introduction}
\label{sec:introduction}
\IEEEPARstart{P}{acket} classification, the process of deciding the fate of an incoming packet based on a set of rules, is a fundamental element of various networking functions, e.g., essential routing policies, network traffic isolation, packet filtering, or even preventing unauthorized access and Denial-of-Service (DoS).
Therefore, packet classification has to be robust, dependable, and highly efficient.
While packet classifiers implemented in hardware easily satisfy these needs, virtual switches deployed within data centers to route traffic between the virtual workloads and the public Internet still face performance challenges~\cite{hypercuts}, and \textit{can} have severe corner cases, especially, as the number of flow rules and matching fields\footnote{Current OpenFlow specification (v1.5~\cite{openflow_spec}) supports more than 40 different header fields a flow rule can match on.} grow~\cite{packet_classification_computational_approach}.

Thus, packet classification algorithms have been studied extensively~\cite{TSS, tuplemerge, efficuts, hierarchical_cuts, harp}. 
Among many approaches, the Tuple Space Search (TSS,~\cite{TSS}) scheme has captivated a wide range of virtualized packet classification systems, e.g., cloud systems~\cite{kubernetes-ovn,openstack-ovn}, stateful NAT implementations~\cite{stateful_nat1}.

Along with several software switch implementations (e.g., VPP~\cite{vpp}, Hyperswitch~\cite{hyperswitch}), Open vSwitch (OVS), the de facto standard of OpenFlow-based SDN switching, uses the TSS scheme for packet classification as well (\cref{sec:backgroud__packet_classification}).
Due to its continuous performance improvements, OVS has become widely used in production cloud environments (e.g., OpenStack~\cite{openstack-ovn}, Kubernetes~\cite{kubernetes-ovn}) for packet switching and firewall implementations for the tenant workloads.

Our recent work~\cite{ovs_dos} has revealed an algorithmic deficiency in the TSS implementation in OVS.
Particularly, we demonstrated a new low-rate DoS attack, called Tuple Space Explosion (TSE), which can quickly degrade the performance of OVS to ${\sim}1\%$ of its baseline with less than $1$ Mbps attack traffic.
In contrast to common DoS attacks, TSE has no specific pattern and only requires arbitrary packets, making the detection and mitigation cumbersome.
TSE only depends on the internal state of the target switch that makes it vulnerable to DoS attacks to randomly-generated inputs.
In~\cite{ovs_dos}, we investigated such system states and showed that the common Access Control Lists (ACLs) given to tenants as default according to the security best practices (e.g., \texttt{Whitelist+DefaultDeny}), are particularly vulnerable to the TSE attack (cf.~\cref{sec:background__tse}).
Besides synthetic measurements with a standalone switch, we have also showed the effectiveness of TSE against Openstack and Kubernetes environments.

However, all investigated scenarios in~\cite{ovs_dos} have one crucial aspect in common.
Open vSwitch was solely using its default kernel datapath for fast packet processing and was installed by the underlying system's own packet manager. 
This implies the following consequences. 
Applications installed in such a way have their kernel modules developed and maintained by a different community rather than the application's developers.
In particular, the kernel module of OVS provided by the vanilla Linux kernel (hereafter, termed as \ovskernel) is maintained by the Linux kernel developers, while OVS installed from source~\cite{ovssourcecode} (henceforth referred to as \ovssource) is managed by the core developers of OVS, which \textit{can} result in non-negligible performance improvements.

On the other hand, despite the continuous performance improvements of the Linux kernel networking stack on x86 architectures~\cite{linux_kernel_networking_perf}, the traffic demands driven by the increasing throughput of commodity networking interfaces require an order of magnitude higher performance, i.e., $100+$ Gbps compared to $10$ Gbps.
Moreover, the packet processing speed we can achieve via the kernel datapath
is highly affected by several other factors (e.g., OS scheduler, multiprocessing capabilities, context-switching, interrupts).
Software acceleration techniques, such as Intel's Data Plane Development Kit (DPDK,~\cite{dpdk}), however, can reduce this overhead on x86 architectures, and software modules, like OVS, can be rewritten to take advantage of it.
Particularly, OVS has already been enhanced with DPDK~\cite{ovs-dpdk} (hereafter, termed as \ovsdpdk) and shown to be able to forward ten times more traffic than before, keeping up with the 
carrier-grade needs.

In this paper, we evaluate the feasibility of the TSE attack against different settings of OVS, i.e., \ovskernel (\cref{sec:evaluation__ovskernel}), \ovssource (\cref{sec:evaluation__ovssource}), and \ovsdpdk (\cref{sec:evaluation__ovsdpdk}). 
Then, we enhance the original TSE attack to accommodate all examined variations where it has shown to be less (or not) effective (at all).
Our contributions can be summarized as follows.
\begin{itemize}
    \item First, we evaluate the latest stable release of \ovskernel and test whether it is still vulnerable; we find that it indeed is vulnerable (\cref{sec:evaluation__ovskernel}).
    Then, we scrutinize \ovssource, i.e., we investigate whether there is any difference (in terms of performance) compared to \ovskernel (\cref{sec:evaluation__ovssource}). 
    Notably, we show that the essential cache implementation (\cref{sec:background__ovs}) of the kernel datapath in \ovssource has more layers than in \ovskernel, which causes non-negligible performance improvement. 
    As the last part of our feasibility study, we evaluate TSE against \ovsdpdk, which, by using the faster DPDK-based user space datapath, is supposed to be less vulnerable; and we confirm this (\cref{sec:evaluation__ovsdpdk}).
    \item After revealing that \ovsdpdk is robust against TSE (\cref{sec:evaluation__ovsdpdk}), we scrutinize its source code to understand the fundamental differences (\cref{sec:mfc_ovsdpdk}) and show that the TSS algorithm itself is significantly enhanced with a ranking process in the tuple space making the considerably higher rate victim flows to be found much faster (\cref{sec:mfc_ovsdpdk__ranking}).
    \item To defeat this ranking process, we propose TSE 2.0.
    In TSE 2.0, by carefully switching the attack on and off, we let some tuples in the tuple space to expire (i.e., disappear) and re-spawn. Thereby, TSE 2.0 keeps the ranking process busy.
    We show that TSE 2.0 is effective and can cause a full denial-of-service attack against \ovsdpdk running on one core (\cref{sec:tse_2.0}).
    Last but not least, we further enhance the improved TSE attack (TSE 2.1) to accommodate multi-core settings, i.e., we scale up \ovsdpdk to multiple CPU cores (\cref{sec:tse_2.1}). 
    We show that as long as OVS is limited to four cores, TSE 2.1 can mount a low-rate denial-of-service attack.
    
    When more than four cores are available, the attack rate required to degrade the performance of OVS to $0$ will not render TSE 2.1 low-rate anymore (\cref{sec:evaluation__methodology_and_threat_model__low_rate}), i.e., it would require more than $24,000$ pps.
\end{itemize}

Note, our aim is to carry out a comprehensive study on TSE (beyond~\cite{ovs_dos}). 
In particular, we shed light on the fact that irrespective of any libraries, datapath implementations or (cloud) environments, as long as TSS and, particularly, OVS is part of a system, it is vulnerable to DoS attacks.

\section{Background}
\label{sec:background}
We first describe the fundamentals of the packet classification in SDN and the widely adopted Tuple Space Search algorithm.
Subsequently, we discuss the essential operation of Open vSwitch (OVS~\cite{ovs}), how its kernel datapath and the implementation of the Tuple Space Search Algorithm~\cite{TSS} look like. 
Finally, we give an overview of the Tuple Space Explosion attack~\cite{ovs_dos}. 

\subsection{Packet Classification in SDN}
\label{sec:backgroud__packet_classification}
The process of categorizing packets according to the flow rules matching on packet header fields is called \textit{packet classification}. 
All packets belonging to the same flow obey the pre-defined flow rule they match on and are processed in the same way. 
These flow rules 
can be (bitwise) masked to be able to match on an arbitrary part of the full packet header solely.
The flow table 
can have hundreds of thousands of flow rules, each matching on arbitrary parts of the header.

Since flow rules can overlap, i.e., a particular packet may match on multiple flow rules simultaneously, priorities are assigned to each of them; the one having higher priority takes precedence. 
Flow tables with priorities and overlapping flow rules are termed as order-dependent tables, while a table without priorities and overlaps is called an order-independent table.
Even if the latter is much simpler~\cite{Kogan:2014:SAX:2619239.2626294}, many software switches (including OVS) support the former due to its flexibility and generality (i.e., wildcarding), allowing complex packet processing logic to be described concisely.

Among the variety of algorithms available (e.g., linear search, set-pruning tries~\cite{set_pruning_tries}, decision trees~\cite{efficuts,harp,hierarchical_cuts}), 
the Tuple Space Search (TSS,~\cite{TSS}) scheme is heavily used in many hypervisor switches as a packet classifier including OVS~\cite{ovs}, VPP~\cite{vpp}, HyperSwitch~\cite{hyperswitch}, etc. 
In a nutshell, TSS collects the flow rules matching on the same set of header bits (e.g., /24 destination prefix) into a hash (i.e., a tuple) in which masked packet headers can be found fast ($\mathcal{O}(1)$~\cite{TSS}).
However, all masks and associated hashes, i.e., the tuple space, is searched sequentially, introducing $\mathcal{O}(n)$ steps in case of $n$ different masks/tuples (see examples later in~\cref{sec:background__tse}).

\subsection{Open vSwitch} 
\label{sec:background__ovs}
OVS is a well-known open-source, multi-layer, production-quality software switch that enables massive network automation through programmatic extensions, while supporting standard management interfaces and protocols, e.g., NetFlow~\cite{Netflow_short}, port mirroring, VLANs. 
It can be operated both as a software switch running within a data center (i.e., as hypervisor switch), and as the control stack for switching silicon. 
Due to its flexibility and generality, not to mention the community support behind it, OVS is heavily used in several production environments such as OpenStack~\cite{openstack-ovn} and Kubernetes~\cite{kubernetes-ovn}.

The quintessential part of OVS, we have to understand, is its packet processing behavior.
The packet processing logic is described at a high level by the order-dependent flow tables (cf.~\cref{sec:backgroud__packet_classification}) running in the user space, referred as \textit{slow path}, due to the packet classification complexity.
In order to fasten the classification process, OVS has introduced flow caches\footnote{In OVS, flow caches are considered the fast path.}; 
once a packet has matched on a flow rule in the flow table, the corresponding information (i.e., packet headers and action) will be cached in the kernel space for future references.
Therefore, the subsequent packets of the same flow will be processed much faster.
Practically speaking, OVS implements two layers of caches; the \textit{Exact Match Cache (EMC)} (also known as the microflow cache) and the \textit{MegaFlow Cache (MFC)}.
In a nutshell, when a packet arrives, it is first looked up in the EMC; if there is a \texttt{cache miss}, the packet header will be looked up in the MFC. 
If there is no \texttt{cache hit} in the MFC either, as a last resort, the packet header will be matched in the flow table\footnote{A packet either explicitly matches on a flow rule, or it is implicitly caught by the default ``drop-all'' rule.} (i.e., in the slow path).

EMC serves as a short-term memory, where the whole header of the packet is cached as it is, i.e., without any wildcard bits.
From an implementation aspect, the data structure is a simple hash table, where look-ups occur fast.
Since each packet belonging to the same flow \textit{may} have different values in the header (e.g., different Time-to-Live values), they all can constitute an individual entry in the EMC.
Thus, for space-efficiency, the size of the EMC is fixed (i.e., 8,192~\cite{emc}); hence, it can be easily saturated during normal operation~\cite{ovs-dpdk-ranking,ovs_dos}. 
Therefore, only the most frequent flows occur in the EMC; others are retained in the MFC, while infrequent flows always have to go through the whole packet processing pipeline.

In contrast to the EMC, the MFC
allows arbitrary bitwise wildcarding.
Introducing MFC to OVS has enabled the most noticeable performance improvement so far~\cite{ovs-ludicruous}.
Due to the wildcards, the TSS scheme used in the slow path has been adopted to the MFC as well; however, TSS in the MFC does not know about flow priorities; all tuples are disjoint to make packet classification simpler,
yet having worst-case exponential complexity.
In particular, as long as the number of masks is kept in a reasonable range (e.g., a couple of hundreds of masks~\cite{ovs_dos}), packet processing is expected to be close to line rate; otherwise, the exhaustive linear search process can significantly slow down the packet processing performance.
This very property is the basis of the TSE attack, i.e., it aims to inflate the tuple space in the MFC to make the linear search process to last as long as possible (see below).

%


\subsection{Tuple Space Explosion~\cite{ovs_dos}: Overview}
\label{sec:background__tse}
In this section, we give an overview of the TSE attack we proposed recently~\cite{ovs_dos}. 
In particular, we discuss how it works, what the requirements are, the threat model considered, and in what environments we carried out the measurements in \cite{ovs_dos}. 
Note, we focus on the operation of the MFC only; we assume the EMC is either switched off or fully saturated by regular traffic, i.e., the fate of a packet is decided by the MFC.

In TSE, an attacker sends a specific packet sequence towards OVS, which, if having the corresponding ACL installed (see later), results in a vast amount of masks in the tuple space; 
thereby slowing down the packet classification to a great extent, practically leading to a Denial-of-Service (DoS)~\cite{ovs_dos}.
Since each tuple in the MFC has a timeout period of 10 seconds, spawning more than a couple of hundreds of tuples each requiring one sole appropriate packet to be received, TSE is considered as a low-rate attack (\cref{sec:evaluation__methodology_and_threat_model__low_rate}).


To show the relationship between the attacker's packet sequence, the flow rules installed, and the different masks in the MFC, we borrow the example from~\cite{ovs_dos}.
Consider a hypothetical 
protocol (\textit{HYP}) and 
the flow table shown in \cref{tab:flow_table}, where wildcards are denoted by \texttt{*}.
There is an exact match \texttt{allow rule} for the packets having \texttt{HYP} header \texttt{001}, 
and a ``catch-all'' \texttt{deny rule} for everything else.

%
\begin{table}
\centering
\vspace{-1em}
\subfloat[Sample Flow Table]
{
\begin{footnotesize}

	\begin{tabular}{|c|c|}
		\hline
		\texttt{HYP}     & \texttt{action} \\
		\hline
		0 0 1  & \texttt{allow}\\
		{*} & \texttt{deny}\\
		\hline
	\end{tabular}
\end{footnotesize}
\label{tab:flow_table}
}
\subfloat[Corresponding MegaFlow Cache.]
{
\begin{footnotesize}
	\begin{tabular}{|c||c|c|c|}
	    \hline
	    {\#} & Key & Mask & Action \\
	    \hline
	    {\#1} & 001 & 111 & \texttt{allow}\\
	    {\#2} & 100 & 100 &  \texttt{deny}\\
	    {\#3} & 010 & 110 &  \texttt{deny}\\
	    {\#4} & 000  & 111 &  \texttt{deny}\\
	    \hline
	\end{tabular}
\end{footnotesize}
\label{tab:mfc}
}
\label{tab:flow_table_and_mfc}
\caption{Sample flow table and the corresponding MFC.}
\end{table}

Now, suppose the switch receives a packet with \textit{HYP} header \texttt{001}. 
Since initially, the MFC is empty, 
the packet is deferred to a full-blown flow table processing, where it will match on the \texttt{allow} rule.
Accordingly, the associated action will be taken, and the new \texttt{(key,mask)} pair \texttt{(001,111)} will be cached into the MFC. 
Since it is an exact-match flow rule in the flow table, it will be an exact match entry (no wildcarded bits) in the MFC as well (row \#1 in \cref{tab:mfc}).

Next, assume a second packet with \textit{HYP} header \texttt{100} is received.
In this case, the lookup process in the MFC 
takes each mask $m \in \mathcal{M}$ one by one, applies $m$ to the header, and looks up the resulting bit-vector in the corresponding hash $H_m$.
If the lookup succeeds, a cache hit is returned; otherwise, it resorts to the next mask. 
Since the lookup fails in this case, 
the packet goes up to the flow table, and accordingly, a new MFC entry will be created.
However, 
due to the wildcarded \texttt{deny rule} 
many further packets not having \texttt{HYP} header \texttt{001} match on, OVS carefully installs a new masked entry in the MFC 
to significantly reducing the overall memory footprint. 
Each new entry has to satisfy three invariants; \textit{(1)} cover the packet itself, which caused its generation, \textit{(2)} be disjoint from all other masked entries currently present in the MFC, and \textit{(3)} be as broad as possible,
i.e., it has to cover as many different packets as possible that would match on the corresponding sole \texttt{deny rule} in the flow table.

Accordingly, when the packet with \textit{HYP} header \texttt{100} is received, the \texttt{(key, mask)} pair in row \#2 in \cref{tab:mfc} will be installed in the MFC.
Observe that since the key itself already differs in the first bit compared to the \texttt{allow rule}, the rest of the bits are not relevant anymore, i.e., 
any further packet with a \texttt{HYP} header that is either \texttt{100}, \texttt{101}, \texttt{110}, or \texttt{111} should be dropped; hence the mask \texttt{100} in row \#2.
For brevity, we do not discuss further how the final MFC in \cref{tab:mfc} is reached, however, row \#3 and \#4 will be spawned similarly by receiving a packet that has a different \texttt{HYP} header than the ones mentioned before.

Nevertheless, observe that for the 3-bit hypothetical \texttt{HYP} header and the simple \texttt{Whitelist+DefaultDeny} ACL, the maximum number of \textit{four} \texttt{(key, mask)} pairs can be created in the MFC by sending only \textit{four} carefully chosen packets; each being covered by one entry in the MFC.
Thus, the number of \texttt{(key, mask)} pairs covering all possible packets significantly increases with the number and bit-width of the headers the installed flow rules match on.
Hence, establishing a logical \texttt{OR} relation between the \texttt{allow} rules on more header fields will create an \texttt{AND} connection on the \texttt{drop} rule.
Therefore, to test each header field \textit{at the same time}, we need to test each combination of \texttt{(key, mask)} pairs for the individual headers resulting in a \textit{multiplicative increase in the tuple space}~\cite{ovs_dos}.
\begin{table}[t!]
\begin{center}
\begin{footnotesize}
 \begin{tabular}{|c|c|c|c||c|}
    \hline
    Rule id & \texttt{ip\_src} & \texttt{tcp\_src} & \texttt{tcp\_dst} &\texttt{action}\\
    \hline
    \#1 & \texttt{*}    & \texttt{*} & \texttt{80} & \texttt{allow}\\
    \#2 & \texttt{10.0.0.1} & \texttt{*}    & \texttt{*}    & \texttt{allow}\\
    \#3 & \texttt{*}    & \texttt{12345} & \texttt{*}  & \texttt{allow}  \\
    \#4 & \texttt{*}    &  \texttt{*} & \texttt{*} & \texttt{deny}\\
    \hline
  \end{tabular}
\end{footnotesize}
\end{center}
\vspace{-1em}
\caption{Simple ACL of a full-blown TSE attack~\cite{ovs_dos}.}
\label{tab:malicious-acl}
\end{table}
This means that a typical ACL matching on the source IP address and Layer-4 source and destination ports (e.g., ACL in Table~\ref{tab:malicious-acl}) can easily result in thousands of MFC masks. 
Consequently, this type of ACLs with an \texttt{OR} relation between the targeted header fields has become the sweet-spot for the TSE attack; \textit{hence the name Tuple Space Explosion}.


\noindent
\subsubsection{Threat model}
\label{sec:background__tse__threatmodel}
In~\cite{ovs_dos}, we considered a simplified cloud infrastructure, which provides APIs for the tenants to configure their ACLs in their logical switches. 
Even though the tenants perceive these virtualized switches as their physical switches, they are only logically separated, and all of them are realized by the same individual software switch instance. 
Therefore, all tenant workloads being scheduled to the same hypervisor inherently share the switching fabric as well (e.g., the MFC).
Hence, if the MFC of a tenant's logical switch is highly populated, it has a severe impact on the packet classification performance for all other co-located tenant workloads. 

We assumed the attacker knows that the target system uses OVS, which has an algorithmic deficiency 
in its MFC. 
The attacker aims to send malicious but legitimately looking packets towards OVS, which will 
cause a DoS to the rest of the tenants. 
Two approaches were devised to the attack, \textit{Co-located TSE}, and \textit{General TSE}. 
In the former, the attacker has a leased resource in the target system, hence, can install specific rules into her logical switch, and send the corresponding packet sequence towards it. 
In the latter, however, the attacker neither has resources nor access to any of the installed ACLs, thereby forced to send random packets instead.
Even though we showed that sending random packets at a rate less than \mbox{$1$ Mbps} can still degrade the victim throughput to almost $10\%$ of its baseline~\cite{ovs_dos}, in this paper, we consider the case of the Co-located TSE \textit{only}; all findings below can be generalized to the General TSE.

%
%

\subsubsection{Environment}
\label{sec:background_tse__environment}

\begin{figure}[t!]
	\vspace{-1em}
	\centering
	\includegraphics[width=\linewidth]{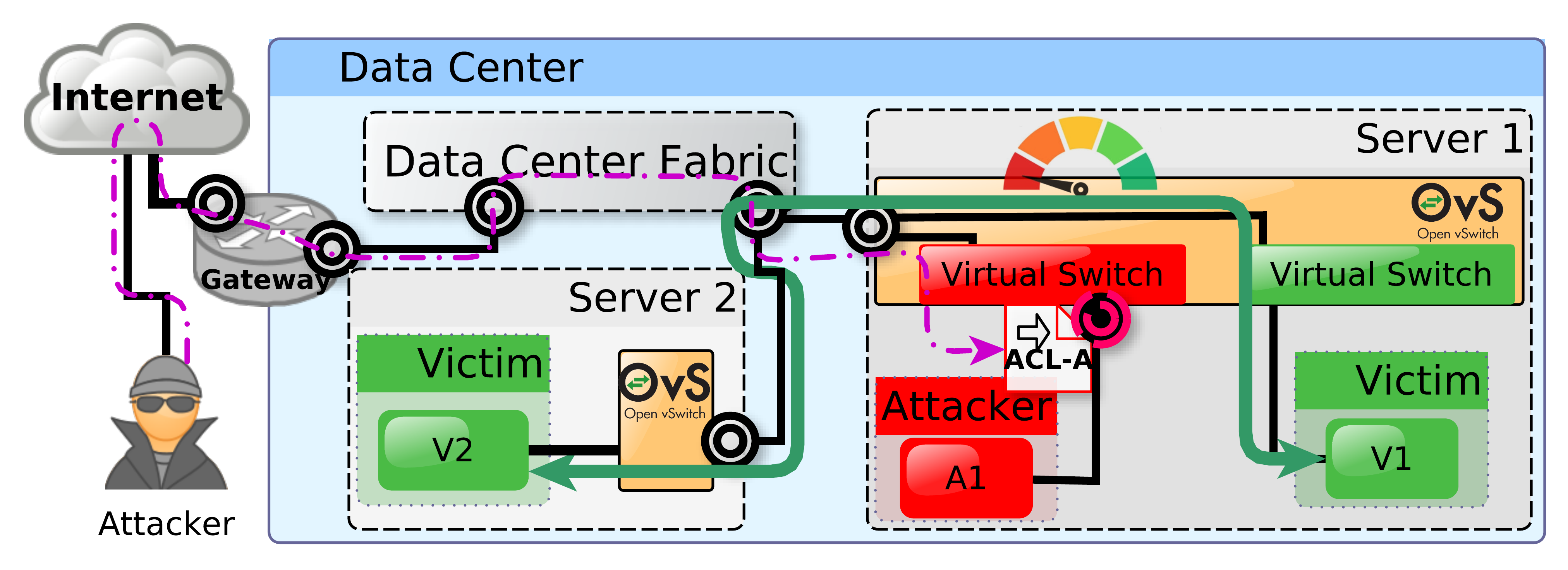}
	\vspace{-2em}
	\caption{Simplified cloud infrastructure used for evaluation.} 
	\label{fig:attack-model}
\end{figure}

First, baseline measurements have been made within a synthetic KVM environment to present that the TSS implementation in OVS is vulnerable to TSE attacks. 
In particular, two servers (Server 1 and Server 2) hosting multiple VMs are connected back to back (cf.~\cref{fig:attack-model}).
Each server, according to the best practices (e.g.,~\cite{hypervisor_switch_best_practice1}, 
runs a single OVS instance as a hypervisor switch to route traffic among the VMs; within and between the servers.
The victim's workload is distributed among the servers; one of its VM is on the first server, while the other is on the second server.
Between them, the victim has a continuous traffic transmission simulating a backend-frontend communication (green arrow in~\cref{fig:attack-model}).

The attacker also has a VM co-located with one of the victim's VM, where she defines her malicious ACL (cf.~\cref{tab:malicious-acl}).
Then, from outside of Server 1, i.e., either from Server 2 (not shown for brevity) or from the Internet (purple dash-dotted arrow), the attacker sends her malicious but legitimate low-rate attack traffic towards her own VM, i.e., to ACL-A.

In~\cite{ovs_dos}, we demonstrated the devastating effects of TSE in such an environment.
We also showed that even if the TSS implementation is offloaded to the hardware with a suitable network interface card, 
the system is still vulnerable.
Finally, we showed the impact of TSE attack in different environments where all management tasks (e.g., workload instantiation, OVS configuration) are done by real, production-quality orchestrators such as OpenStack and a Kubernetes.

\subsubsection{Limitations}
\label{sec:background__tse__limitations}
Even though the impact of TSE was the same irrespective of the environment, all scenarios had one crucial aspect in common; 
OVS was installed using the underlying system's packet manager\footnote{E.g., via \texttt{apt-get install openvswitch-common}.}.

OVS and its kernel datapath comprising the caching layers became part of Linux in 2012 with the Linux 3.3 kernel debut release~\cite{ovs_in_linux}.
While changes to OVS are pushed to the mainline upstream kernel (i.e., available in \ovskernel), OVS has an up-to-date ``out-of-tree'' kernel module maintained by the OVS community (i.e., the case of \ovssource), from which the new releases are (selectively) pulled up (for \ovskernel) by the Linux kernel developer(s).
This renders \ovskernel, the variant we evaluated \textit{exclusively} in~\cite{ovs_dos}, to be different from \ovssource.
During the discussion of~\cite{ovs_dos}, 
it turned out that this difference not only lags behind the latest version of \ovssource, but some features \textit{may} be implemented differently.
For instance, due to \textit{not} being in favor of exact match caching in the kernel space, EMC is one of the features that has been entirely disabled in \ovskernel; however, it is a fundamental part of \ovssource.

Furthermore, we did not scrutinize \ovsdpdk in~\cite{ovs_dos}, which can provide ten times higher performance, and in addition, the linear search process of the packet classification in the MFC is enhanced~\cite{ovs-dpdk-ranking}.
In particular, the developers have introduced a ranking process to the tuple space (cf.~\cref{sec:mfc_ovsdpdk__ranking}), which is conducive to the linear search process in TSS, to find a corresponding tuple much faster; the higher number of hits a tuple has, the faster it will be found.
This can result in the case that the tuples spawned by the low rate TSE attack have no significant impact on the overall lookup process as the benign flow(s) have orders of magnitudes higher hits for their corresponding tuples.
Therefore, the lookup process would only require more time for the TSE's attack traffic itself; however, due to TSE being low rate, the overall impact is negligible (we discuss the details in \cref{sec:evaluation__ovsdpdk}).

Therefore, in this paper, we aim to evaluate and improve TSE against different OVS variants.




\section{Feasibility of the TSE Attack}
\label{sec:evaluation}
In~\cite{ovs_dos}, \ovskernel was analyzed exclusively irrespective of the environments.
Therefore, to comprehensively study the feasibility of the TSE attack, we define three scenarios where OVS is not only more up-to-date but all possible ways of implementing its datapath are considered, which could have significant performance differences.

\subsection{Scenarios}
\label{sec:evaluation__scenarios}

\noindent
\textbf{\ovskernel. } 
The TSE attack was initially discovered in the standalone OVS v2.9.2 (v2.7 in Kubernetes and the unstable v2.9.90 in OpenStack) using \ovskernel for their datapaths.
Therefore, first, we investigate whether the TSE attack is still feasible in a similar environment but used against a more up-to-date version of OVS (2.10.0\footnote{i.e., \texttt{2.10.0+2018.08.28+git.8ca7c82b7d+ds1-12+deb10u1}}) available in recent stable Debian 10.4 systems. 

\noindent
\textbf{\ovssource. } 
We scrutinize the performance of \ovssource, i.e., we install and configure OVS from its source~\cite{ovssourcecode}, and evaluate whether it is affected by the TSE attack the same way; \textit{it is not}.

\noindent
\textbf{\ovsdpdk. } 
Besides the kernel datapath, OVS can be compiled with Intel's DPDK userspace libraries~\cite{ovs-dpdk} that specifically tailored to accelerate packet processing workloads on a wide variety of CPU architectures.
It has been shown in several studies (e.g.,~\cite{ovs-dpdk-perf,ovs-dpdk-performance-yahoo}) 
that DPDK can scale OVS performance significantly (even ten times more~\cite{ovs-dpdk}), especially when multiple CPU cores are exclusively assigned for packet processing.
Therefore, we evaluate whether the TSE attack 
can deteriorate the performance of \ovsdpdk. 

By experimental evaluations, we show that the TSE attack does not perform equally well in all examined environments; hence we take a closer look at the possible reasons.
In particular, we show that despite our findings in~\cite{ovs_dos}, the use of EMC in \ovssource significantly alleviates the impact of the TSE attack, which requires orders of magnitude higher attack rate to achieve the same goal.

Furthermore, we show that the optimizations (e.g., tuple ranking~\cref{sec:background__tse__limitations}) made to the operation of the MFC in the DPDK datapath renders TSE attack much less effective.
Accordingly, in a subsequent section (\cref{sec:mfc_ovsdpdk}), we deeply analyze the operation of the DPDK datapath and carry out several modifications to the original TSE attack to accommodate these optimizations.
In particular, we propose TSE 2.0 (\cref{sec:tse_2.0}) and TSE 2.1 (\cref{sec:tse_2.1}), two effective and low-rate attacks against the most prominent DPDK-based software packet classifier, \ovsdpdk, running on one or multiple cores, respectively. 

\noindent
\textbf{Mitigation. }
In this paper, we aim to improve the original TSE attack to be effective against multiple enhanced implementations of the OVS datapath.
However, all our enhanced variants are exploiting the same algorithmic deficiency as discovered in~\cite{ovs_dos}. 
Consequently, since the mitigation approach proposed in~\cite{ovs_dos} is still effective against all our proposals, discussion of further mitigation solutions is out of the scope of this paper.

\subsection{Threat Model and Methodology}
\label{sec:evaluation__methodology_and_threat_model}
As mentioned in~\cref{sec:background__tse__threatmodel}, we focus on and reproduce the Co-located TSE attack only. 
Furthermore, as the original TSE attack was equally successful in multiple environments, 
we carry out our feasibility study in the synthetic KVM environment exclusively.
\cref{tab:hw_sw_details} lists the hardware and software configurations on top of Debian 10.4 we used in our analysis.

Our threat model is completely in line with the original one (cf.~\cref{sec:background__tse__threatmodel} and \cref{fig:attack-model}), and we use the same ACL the attacker is targeting (cf.~\cref{tab:malicious-acl}).
Last but not least, similarly to~\cite{ovs_dos}, we consider three different attack use cases w.r.t. the different header field combinations it exploits.
\begin{enumerate}
    \item \textbf{\texttt{DP}} use case is strictly against the L4 destination port, i.e, against \texttt{allow rule \#1}, and the \texttt{deny rule \#4} in~\cref{tab:malicious-acl}. 
    This enables the attacker to spawn 16 different tuples in the MFC. 
    Due to the MFC timeout period of 10 seconds (\cref{sec:background__tse}), this attack has the lowest rate, and requires only $16$ packets to send within $10$ seconds to be successful, resulting in ${\sim}1$ kbps attack rate (considering the smallest possible packet size of 64-byte packets).
    \item \textbf{\texttt{SP{\_}DP}} use case is against the source and destination ports, i.e, against \texttt{allow rules \#1, and \#2}, and \texttt{deny rule \#4} in \cref{tab:malicious-acl}. 
    This leads to 16*17 = 272 tuples in the MFC. 
    According to the MFC timeout, the required attack rate is less than $30$ packets per second, i.e., $<20$ kbps (considering 64-byte packets).
    \item \textbf{\texttt{SIP{\_}SP{\_}DP}} is the full-blown TSE attack. It is against the source IP and source and destination ports, i.e., against the whole ACL in~\cref{tab:malicious-acl}. 
    This allows the attacker to spawn as a high number of tuples as ${\sim}9000$ in the MFC by not exceeding a still very low attack rate of $605$ kbps (considering 64-byte packets). 
\end{enumerate}

Note that the used packet traces (generated according to~\cite{ovs_dos}) contain slightly more packets than the number of masks they can spawn due to the cross-product of all packets that match on all the corresponding header fields.
Therefore, some packets share the same masks (for instance, recall \textit{HYP} header and the packets \texttt{001} and \texttt{000} in~\cref{sec:background__tse}).
In particular, the \texttt{DP}, \texttt{SP\_DP} and \texttt{SIP\_SP\_DP} trace contains $17$ packets ($1$ packet more), $289$ packets ($1\times17$ packets more), and $9.537$ packets ($1\times17\times33=561$ packets more).

During our measurements, for the first $20$ seconds, OVS only processes benign traffic, and besides the tenant workloads, no additional task is run on the servers. 
Furthermore, there is a continuous \texttt{iperf} session between the victim's two VMs located at each server, and the measured throughput (reported by \texttt{iperf}) is observed.
Then, after the $20^{\text{th}}$ second, the attack is launched. 
To ease the calculations, we use $100$ pps attack rate in the \texttt{DP} and \texttt{SP{\_}DP} use cases, while $1000$ pps in the \texttt{SIP{\_}SP{\_}DP} to saturate the MFC. 
 
\begin{table}[t!]
\begin{center}
\begin{scriptsize}
\begin{tabular}{|c|c|}
\hline
    Property &  Value\\
    \hline
    CPU \& Memory & Intel(R) Xeon(R) Gold 6230 CPU \& 93G\\
    NIC & Mellanox CX-5\\
    \ovskernel & 2.10 (kernel version 4.19.0-8-amd64) \\
    \ovssource & 2.13.90\\
    \ovsdpdk & 19.11\\
    \hline
\end{tabular}
\end{scriptsize}
\caption{HW/SW versions of our testbed}
\label{tab:hw_sw_details}
\end{center}
\end{table}

\subsubsection{Low-rate Attack}
\label{sec:evaluation__methodology_and_threat_model__low_rate}
In general, an attack is considered low rate, when it is challenging to be accurately detected by standard firewalls or anomaly-based detection schemes due to its low-rate nature~\cite{low_rate_dos_tifs1,low_rate_dos_tifs2,low_rate_dos_tifs3} 
For instance, TCP-targeted DoS attacks are one of the most common low-rate DoS attacks.
Here, the attack traffic is divided into packet burst and it is considered low-rate if within each attack period $T$ with burst rate $R$, the burst length $L$ is much smaller than $T$, i.e., the average attack rate $R \times \tfrac{L}{T}$ is relatively small.

On the other hand, we can consider an attack to be low-rate if it does not comprise more bandwidth than, say, $10-20$\% of the background traffic~\cite{low-rate-dos_percentage}.

Here, we rather focus on a simplified yet more explicit property of the attack, which does not (proportionally) depend on any other factor mentioned above.
Thus, we consider an attack to be low rate if it does not send more than $15,000$ packets per second, which, when the smallest packet size of 64 bytes is considered, equals to ${\sim}10Mbps$ bandwidth.

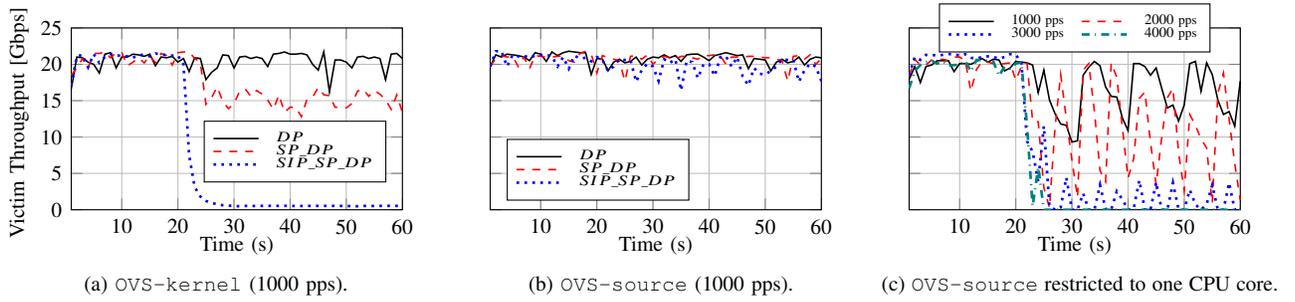
\begin{figure*}[t!]
\vspace{-1.5em}
\centering
\subfloat[\ovskernel (1000 pps).]
{
\begin{tikzpicture}
\begin{axis}[
	width=.33\textwidth,
	height=4cm,
    ymax=25,
    xmin=1, 
    xmax=60,
    ymin=0,
    xtick={0,10,20,30,40,50,60},
    ytick={0,5,10,15,20,25},
    xlabel={Time (s)},
    ylabel={Victim Throughput [Gbps]},
    legend cell align=left,
    mark=yes,
	mark size = 0pt,
    legend style={
            at={(.4,.15)},
            anchor=south west,
            column sep=0.15cm,
			row sep=-0.15cm,
            legend columns=1,
            font=\tiny,
    },
    ymajorgrids=true,
    xmajorgrids=true,
    grid style=solid,
    every tick label/.append style={
    	font=\footnotesize
    },
    every axis y label/.style={
    	at={(-0.1,0.48)},
    	font=\footnotesize,
    	anchor=south,
    	rotate=90,
    },
    every axis x label/.style={
    	at={(0.5,-0.3)},
    	anchor=south,
    	font=\footnotesize,
    },
    ]
\addplot [black, solid, line width=.6]  
table[x=time, y=throughput_DP, col sep=comma, mark=none]{figs/OVS_Kernel_Linux.csv};
\addlegendentry{$DP$}

\addplot [red, dashed, line width=.6]  
table[x=time, y=throughput_SP_DP, col sep=comma, mark=none]{figs/OVS_Kernel_Linux.csv};
\addlegendentry{$SP\_DP$}

\addplot [blue, dotted, line width=1]  
table[x=time, y=throughput_SIP_SP_DP, col sep=comma, mark=none]{figs/OVS_Kernel_Linux.csv};
\addlegendentry{$SIP\_SP\_DP$}

\end{axis}
\end{tikzpicture}
\label{fig:tse_against_ovskernel}
}
\hspace{.5em}
\subfloat[\ovssource (1000 pps).]
{

\begin{tikzpicture}
	\begin{axis}[
		width=.33\textwidth,
		height=4cm,
		ymax=25,
		xmin=1,
		xmax=60,
		ymin=0,
		xtick={0,10,20,30,40,50,60},
		xlabel={Time (s)},
		ytick={0,5,10,15,20,25},
		yticklabels={},
		legend pos=south west,
		legend cell align=left,
		mark=yes,
		mark size = 0pt,
		legend style={
			at={(.05,.05)},
			anchor=south west,
			column sep=0.15cm,
			row sep=-0.15cm,
			legend columns=1,
			font=\tiny,
		},
		ymajorgrids=true,
		xmajorgrids=true,
		grid style=solid,
		every tick label/.append style={
			font=\footnotesize
		},
		every axis y label/.style={
			at={(-0.1,0.48)},
			font=\footnotesize,
			anchor=south,
			rotate=90,
		},
		every axis x label/.style={
			at={(0.5,-0.3)},
			anchor=south,
			font=\footnotesize,
		},
		]
		\addplot [black, solid, line width=.6]  
		table[x=time, y=throughput_DP, col sep=comma, mark=none]{figs/OVS_Kernel_Source_code_all_cores.csv};
		\addlegendentry{$DP$}

		\addplot [red, dashed, line width=.6] 
		table[x=time, y=throughput_SP_DP, col sep=comma, mark=none]{figs/OVS_Kernel_Source_code_all_cores.csv};
		\addlegendentry{$SP\_DP$}
		
		\addplot [blue, dotted, line width=1]  
		table[x=time, y=throughput_SIP_SP_DP, col sep=comma, mark=none]{figs/OVS_Kernel_Source_code_all_cores.csv};
		\addlegendentry{$SIP\_SP\_DP$}
		
	\end{axis}
\end{tikzpicture}
\label{fig:tse_against_ovssource}
}
\hspace{.5em}
\subfloat[\ovssource restricted to one CPU core.]
{
	
	\begin{tikzpicture}
		\begin{axis}[
			width=.33\textwidth,
			height=4cm,			
			ymax=25,
			xmin=1,
			xmax=60,
			ymin=0,
			xtick={0,10,20,30,40,50,60},
			xlabel={Time (s)},
			ytick={0,5,10,15,20,25},
			yticklabels={},
			legend cell align=left,
			legend style={
				at={(.5,1)},
				anchor=center,
				column sep=0.15cm,
				row sep=-0.15cm,
				legend columns=2,
				font=\tiny,
			},
			ymajorgrids=true,
			xmajorgrids=true,
			grid style=solid,
			every tick label/.append style={
				font=\footnotesize
			},
			every axis y label/.style={
				at={(-0.1,0.48)},
				font=\footnotesize,
				anchor=south,
				rotate=90,
			},
			every axis x label/.style={
				at={(0.5,-0.3)},
				anchor=south,
				font=\footnotesize,
			},
			]
			\addplot [black, solid, line width=.6] 
			table[x=seconds, y=1000pps, col sep=comma, mark=none]{figs/OVS_Kernel_Source_code_1_core.csv};
			\addlegendentry{$1000$ pps}
			
			\addplot [red, dashed, line width=.6]  
			table[x=seconds, y=2000pps, col sep=comma, mark=none]{figs/OVS_Kernel_Source_code_1_core.csv};
			\addlegendentry{$2000$ pps}
			
			\addplot [blue, dotted, line width=1] 
			table[x=seconds, y=3000pps, col sep=comma, mark=none]{figs/OVS_Kernel_Source_code_1_core.csv};
			\addlegendentry{$3000$ pps}
			
			\addplot [teal, dashdotted, line width=1] 
			table[x=seconds, y=4000pps, col sep=comma, mark=none,]{figs/OVS_Kernel_Source_code_1_core.csv};
			\addlegendentry{$4000$ pps}
			
		\end{axis}
	\end{tikzpicture}
	\label{fig:tse_against_ovssource__one_core_only}
}

\caption{TSE attack with different attack rates against different kernel datapaths (a)(b) with resource constraints (c).}
\label{fig:tse_agains_kernel_datapaths}
\end{figure*}

\subsection{Scenario \ovskernel}
\label{sec:evaluation__ovskernel}
As mentioned before and detailed in \cref{tab:hw_sw_details}, here we use OVS with the version coming from the recent stable Debian repository (10.4) with a kernel version of 4.19.0-8-amd64.
Recall, in this case, the kernel datapath is coming from a different source maintained by a different community\footnote{The related source files (e.g., \texttt{datapath.c, flow.c}) are missing due to being part of the kernel source itself.}, and the exact match cache is not part of the datapath\footnote{The related \texttt{dpif-netdev.h/c} sources are missing.}. 

We perform the attack for all three use cases mentioned above, namely \texttt{DP}, \texttt{SP\_DP} and \texttt{SIP\_SP\_DP}. 
The results are depicted in Fig.~\ref{fig:tse_against_ovskernel}, where the attack starts at the $20^{\text{th}}$ second.
The small spikes before (and after the attack) are the measurement errors of \texttt{iperf} itself.
We have run several experiments, and these minor glitches are always present. 

We observe that the attack works as expected; however, it is not as visible as in~\cite{ovs_dos}, especially for the \texttt{DP} case.
The victim throughput is degraded by more than $50\%$ in~\cite{ovs_dos}, whereas no significant degradation can be seen in our system for the same \texttt{DP} use case.
This can be attributed to fresher OVS version and kernel datapath, and the available resources our system has compared to the test-bed used in~\cite{ovs_dos} (i.e., more up-to-date kernel version of 4.19 compared 4.13, $3900$ Mhz compared to $3200$ Mhz, 93 GB memory compared to 64 GB); the linear search process is not significantly affected by having only $17$ tuples to loop through in the worst case.

In the case of the \texttt{SP\_DP} use case, compared to the $90\%$ drop in~\cite{ovs_dos}, in our system, the victim throughput is only reduced by $25\%$ of the nominal throughout of $20$ Gbps.

Last but not least, in the \texttt{SIP\_SP\_DP} use case, the TSE attacks achieves the complete DoS in our system. 

We can conclude that while having more resources can alleviate the impact of TSE when the number of tuples spawned is not exceeding a few hundreds, when thousands of them are present in the system, the discovered algorithmic deficiency of the TSS becomes clearly visible.

\subsection{Scenario \ovssource}
\label{sec:evaluation__ovssource}
Next, we analyze the TSE attack against \ovssource.
To this end, we obtain OVS from its source code~\cite{ovssourcecode} and compile it with its kernel datapath. 
At the time of writing, the version of OVS we built is 2.13.90, released in September 2019.
Following our methodology above, we perform the same experiments with the three use cases to evaluate \ovssource. 

From the results depicted in Fig.~\ref{fig:tse_against_ovssource}, we observe that the TSE attack is hardly affecting the victim throughput irrespective of the use case.
We also observe that \texttt{iperf} reports more consistent throughput, i.e., the spikes caused by measurement errors are smaller before the attack.
While the throughput starts to be varying more after the attack started, on average, the throughput is comparable with the baseline performance before the attack.

\subsubsection{Restricting \ovssource to one core}
\label{sec:evaluation__ovssource__less_resources}
In order to pinpoint what prevents TSE attack from achieving any degradation against OVS, first, we make \ovssource to run only on a single-core, thereby exclude the available resource from the influential factors.
While curtailing OVS from scaling to multiple cores with DPDK (i.e., \ovsdpdk) is straightforward, the same does not apply to the kernel datapath. 
For instance, the easiest way to restrict OVS to use one core only is to run the whole system in a single-core setting by disabling multiprocessing (via a GRUB \texttt{cmdline} argument).
However, this approach cannot fit our purpose since having two VMs running on top of the base operating system with one core already affects the overall performance significantly if not disabling to run a VM at all.

One might consider other well-known approaches, such as \textit{(i)} the priority-based command-line utilities \texttt{nice, cgroups}, or \textit{(ii)} the CPU-affinity-based \texttt{taskset}.
In the case of \textit{(i)}, even though \texttt{cgroups} (prominently used in container-based environments~\cite{docker_vs_lxc}) provides more granularity than \texttt{nice}, their overall behavior significantly depends on the competing tasks; if there is none (like in our case), OVS can still scale up according to the actual load.

%

On the other hand, \textit{(ii)} \texttt{taskset} can set the CPU affinity of a running process and bind it to a specific CPU core, we observe that only the started process can be restricted, and the later sub-processes OVS instantiates, i.e., \texttt{ksoftirqd} (see details below), do not inherit the \texttt{taskset} setting rendering it infeasible for resource restrictions in case of OVS.


In a nutshell, when using the kernel datapath, the Network Interface Card (NIC) and the operating system handles incoming data via interrupts.
The NIC issues a hard interrupt when a packet arrives, and since hard interrupts are generally CPU extensive, the rest of the packet handling is managed by soft interrupts (softirqs).
In Linux systems, the \texttt{ksoftirqd} is a queuing daemon for all soft interrupt requests that manages interrupts when the system is under a heavy soft interrupt load, i.e., when thousands of packets are received within a second many \texttt{ksoftirqd} processes run on many cores concurrently.

By adopting the findings of~\cite{ksoftirqd}, we limit the number of concurrent \texttt{ksoftirqd} processes, i.e., we implicitly limit the number of cores assigned for packet processing. 

After modifying our system settings, we evaluate \ovssource (now restricted to use only one CPU core) against the TSE attack.
First, we set the attack rate to the default $1000$ pps. 
Recall, we only evaluate the \texttt{SIP\_SP\_DP} use case to achieve the highest possible impact of the TSE attack.
The result, denoted by a black solid line, is shown in \cref{fig:tse_against_ovssource__one_core_only}. 
We observe that after the attack is launched, the victim throughput drops down by ${\sim}45\%$; however, right after it reaches its lowest performance, it rises back to its baseline (i.e., back to $20$ Gbps).
Then, again, after a couple of seconds, the throughput falls back to ${\sim}45\%$, and this sinusoidal behavior repeats until the end of the measurement.

This phenomenon cannot solely be attributed to the presence of EMC in \ovssource as the EMC's capacity is just around ${\sim}9000$~\cite{ovs-dpdk-ranking}, i.e., around the number of different packets the used \texttt{SIP\_SP\_DP} trace has.
This means that while populating the EMC at the beginning (i.e., when the attack is started) \textit{might} cause extra overhead in OVS, after EMC caches all possible packet headers (assuming that the number of hash collisions is negligible), the performance should be consistent.
However, we observe somewhat contrary, particularly when the EMC would be re-populated by other traffic at the same time.
In particular, while the attack traffic populates the EMC, right after processing it, some other benign traffic, say, destined to another workload, evicts all previous entries from the EMC requiring the malicious traffic to be processed in the MFC.
Afterward, the corresponding entries will be cached in the EMC again, and this ``cat-and-mouse'' game continues.

To verify such behavior, we spawn two additional VMs (on the two separate servers), say, for another tenant in the system, and refine the ACL installed in OVS accordingly (i.e., we extend the ACL in \cref{tab:malicious-acl} with two new exact-match flow rules to provide the connection between these two VMs).
Then, we create a similar trace to \texttt{SIP\_SP\_DP} with completely different un-important headers (i.e., TTL values) than the original \texttt{SIP\_SP\_DP} trace has, and replay this new trace 
to let it fully accommodate the whole EMC in the beginning. 
Afterward, we continue with our analysis, i.e., the victim initiates an \texttt{iperf3} session between the two workloads, and in the $20^{\text{th}}$ second, the attacker launches her attack.
We observe the same behavior (not shown for brevity) as expected, however, similar to further unknown side-effects of EMC (e.g., ~\cite{less_throughput_with_emc}), we do not find any reasonable explanation about the same phenomenon in the original setting.

Therefore, as the performance of the lookup process not only depends on the number of masks but the number of packets has to be classified within a second, we iteratively increase the attack rate to study to what extent the attack rate should be increased to achieve a full degradation of the victim throughput.
The degraded throughput of the victim can be seen in Fig.~\ref{fig:tse_against_ovssource__one_core_only} when the attack rate is set to $2000$ (red dashed line), $3000$ (blue dotted line), and $4000$ (teal dash-dotted line) pps, respectively.
As expected, the higher the attack rate, the more influential the attack is, i.e., the more the victim throughput is degraded.
In particular, for $2000$  and $3000$ pps, the throughput of the victim drops down by $45\%$ and $87\%$, respectively.
However, in all cases, a similar tendency can be observed w.r.t. the sinusoidal pattern in case of the attack rate of $1000$ pps, i.e., once the throughput of the victim fully degrades to its minimum, it instantly goes back to its baseline, and this pattern repeats.
On the other hand, the results also confirm that the TSE attack with an attack rate of $4000$ pps we can achieve a complete denial-of-service when \ovssource is restricted to one core only (in our test-bed).

\subsection{Scenario \ovsdpdk}
\label{sec:evaluation__ovsdpdk}
As mentioned in~\cref{sec:background}, compiling OVS with DPDK and having a datapath running in the user space can increase the attainable throughput tenfold~\cite{ovs-dpdk-performance-yahoo,ovs-dpdk-perf}. 
Thus, we compiled and configured OVS to leverage the DPDK libraries.
For the physical connectivity, we relied on the regular DPDK-capable \texttt{mlx5} drivers, while we used the \texttt{dpdk-vhost} subsystem to connect the VMs to the OVS instance.
Since with DPDK, we explicitly have to set how many cores our switch can scale up to, if not stated otherwise, in every measurement, we set OVS to run on a single core only.
Note, in today's systems, a single core can easily accommodate more than 15-20 Gpbs traffic with the smallest packet size of 64-bytes~\cite{dpdk-performance}.

\begin{figure}[t!]
\centering
\vspace{-1em}
\subfloat[TSE attack against OVS-DPDK.]
{
\hspace{-1em}
\begin{tikzpicture}
\begin{axis}[
	height=4cm,
	width=.6\linewidth,
    ymax=20,
    xmin=1,
    xmax=60,
    ymin=0,
    xlabel={Time (s)},
    ylabel={Victim Throughput [Gbps]},
	xtick={0,10,20,30,40,50,60},
    ytick={0,5,10,15,20},
    yticklabels={0,5,10,15,20},
    mark=yes,
	mark size = 0pt,
    legend style={
            at={(.05,.8)},
            anchor=south west,
            column sep=0.15cm,
            row sep=-0.15cm,
            legend columns=1,
            font=\tiny,
    },
	ymajorgrids=true,
	xmajorgrids=true,
	grid style=solid,
	every tick label/.append style={
		font=\footnotesize
	},
	every axis y label/.style={
		at={(-0.1,0.48)},
		font=\footnotesize,
		anchor=south,
		rotate=90,
	},
	every axis x label/.style={
		at={(0.5,-0.35)},
		anchor=south,
		font=\footnotesize,
	},
]

\addplot [black, solid, line width=.6]  
table[x=time, y=throughput_DP, col sep=comma, mark=none]{figs/DPDK.csv};
\addlegendentry{$DP$}

\addplot [red, dashed, line width=.6] 
table[x=time, y=throughput_SP_DP, col sep=comma, mark=none]{figs/DPDK.csv};
\addlegendentry{$SP\_DP$}

\addplot [blue, dotted, line width=1]  
table[x=time, y=throughput_SIP_SP_DP, col sep=comma, mark=none]{figs/DPDK.csv};
\addlegendentry{$SIP\_SP\_DP$}

\end{axis}
\end{tikzpicture}
\label{fig:tse_against_ovsdpdk}
}
\subfloat[Justification of the resurgence.]
{
	\hspace{-2em}
\begin{tikzpicture}[spy using outlines={rectangle, magnification=4, connect spies}]
	\begin{axis}[    
		height=4cm,
		width=.6\linewidth,
		ymax=20,
		xmin=1,
		xmax=60,
		ymin=0,
		xlabel={Time (s)},
		ytick={0,5,10,15,20},
		yticklabels={},
		xtick={0,10,20,30,40,50,60},
		mark=yes,
		mark size = 0pt,
		legend style={
			at={(.55,.95)},
		    anchor=center,
			column sep=0.15cm,
			row sep=-.15cm,
			legend columns=2,
			font=\tiny,
		},
		ymajorgrids=true,
		xmajorgrids=true,
		grid style=solid,
		every tick label/.append style={
			font=\footnotesize
		},
		every axis x label/.style={
			at={(0.5,-0.35)},
			anchor=south,
			font=\footnotesize,
		},
		]

		\addplot [black, solid, line width=.6]  
		table[x=seconds, y=Run 1, col sep=comma, mark=none]{figs/Resurgence_confirmation.csv};
		\addlegendentry{Meas. \#1}
		
		\addplot [red, dashed, line width=.6] 
		table[x=seconds, y=Run 2, col sep=comma, mark=none]{figs/Resurgence_confirmation.csv};
		\addlegendentry{Meas. \#2}
		
		\addplot [blue, dotted, line width=.6]  
		table[x=seconds, y=Run 3, col sep=comma, mark=none]{figs/Resurgence_confirmation.csv};
		\addlegendentry{Meas. \#3}
		
		\addplot [teal, dashdotted, line width=.6]  
		table[x=seconds, y=Run 4, col sep=comma, mark=none]{figs/Resurgence_confirmation.csv};
		\addlegendentry{Meas. \#4}
		
		\draw [lightgray, densely dotted, line width=.6] (21,0) -- (21,20);
		\draw [lightgray, densely dotted, line width=.6] (22,0) -- (22,20);
		\draw [lightgray, densely dotted, line width=.6] (23,0) -- (23,20);
		\draw [lightgray, densely dotted, line width=.6] (24,0) -- (24,20);
		\draw [lightgray, densely dotted, line width=.6] (25,0) -- (25,20);
		\draw [lightgray, densely dotted, line width=.6] (26,0) -- (26,20);
		\draw [lightgray, densely dotted, line width=.6] (27,0) -- (27,20);
		\draw [lightgray, densely dotted, line width=.6] (28,0) -- (28,20);
		\draw [lightgray, densely dotted, line width=.6] (29,0) -- (29,20);
		\draw [lightgray, densely dotted, line width=.6] (30,0) -- (30,20);

		\coordinate (spypoint) at (axis cs:24.5,1);
		\coordinate (magnifyglass) at (axis cs:45,10.5);
		
	\end{axis}
	\spy [blue, size=1.3cm] on (spypoint) in node[fill=white] at (magnifyglass);
\end{tikzpicture}

\label{fig:resurgence_confirmation}
}
\caption{TSE 1.0 attack against \ovsdpdk.}
\label{fig:tse_against_ovsdpdk_and_resurgence}
\end{figure}
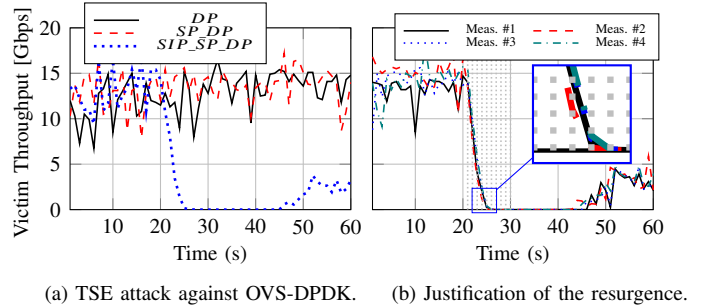

The results for the TSE attack is depicted in Fig.~\ref{fig:tse_against_ovsdpdk}.
The first conspicuous observation is that the average victim throughput before the attack is lower than in the case of OVS with kernel datapath. 
This is attributed to the fact that OVS is properly pinned to a single core only, while the kernel datapath could dynamically scale up.
The second observation is that the sinusoidal effect in both cases, i.e., before and during the attack, is more visible.
This happened due to the fact the inside the VMs, for consistency, we are still using \texttt{iperf3} for throughput measurements, which does not benefit the DPDK-based \texttt{vhost} ports but introduces context switches. 
Note, this is a normal behavior when stress-testing such a system.
Nevertheless, for confirming, we measured the throughput with DPDK \texttt{pktgen} applications running inside the VMs, and the performance was more stable and topped around 15 Gbps.

The performance gain of DPDK can be easily seen in the case of \texttt{DP} and \texttt{SP\_DP}, i.e., the considerably lower number of masks generated (magnitude of hundreds at most) does not impose a barrier for \ovsdpdk.
On the other hand, in the case of \texttt{SIP\_SP\_DP}, the TSE attack is as effective as before; we can observe a complete denial-of-service.

Interestingly, however, the DoS time frame does not last long; TSE is only working for a certain period, i.e., until the ${\sim}45^{\text{th}}$ second.
Subsequently, a resurgence in the victim throughput is observed, although still having a significant degradation level.
In particular, victim throughput goes back to around $2.5-3.5$ Gbps ($\tfrac{1}{5}$ of its throughput) rendering TSE not to be a complete DoS attack but only a degradation-of-service attack.
In the following section, we scrutinize this resurgence and pinpoint the root cause of it.

\section{Analyzing MFC in \ovsdpdk}
\label{sec:mfc_ovsdpdk}
In this section, we first analyze the root cause of the resurgence we observed in the victim traffic in~\cref{sec:evaluation__ovsdpdk}.
We make a deep dive into the DPDK datapath source code and discuss the enhancements made to TSS.
In particular, we note that a ranking has been introduced to the tuple space according to the number of hits, which significantly fastens the linear search algorithms. 
Furthermore, we shed light on the fact that to further improve the packet processing of the most recent flows, all tuples when generated are ranked top.
Then, by exploiting this behavior, we extend the original TSE attack (henceforth referred to as TSE 1.0) to keep the ranking process busy, thereby not letting the victim traffic resurge. 


\subsection{Ranking in the Tuple Space}
\label{sec:mfc_ovsdpdk__ranking}
In 2016, a patch was introduced to \ovsdpdk~\cite{ovs-dpdk-ranking} to solve the possible performance degradation when there are too many tuples by adding a mechanism to sorts the tuples according to the overall number of hits their entries have.
Therefore, whenever a packet of a frequent flow has to be classified, its corresponding tuple will be ranked higher making the linear search process faster to find it.
That is, the corresponding tuple will be relocated up to the front of the tuple space, where the linear search process starts its iteration.
By this new mechanism, the most frequent flows can even achieve $\mathcal{O}(1)$ sub-table lookup only to be classified, which increases the forwarding performance on average by ${\sim}30\%$~\cite{ovs-dpdk-ranking-patch}.
Hereafter, we use the terms sub-table, mask, and tuple \textit{interchangeably}.

Next, we take a closer look at the implementation. 

\subsubsection{Source Code Analysis}
\label{sec:mfc_ovsdpdk__ranking__source_code}
The relevant source code for the Megaflow Cache and the classifier for the user space datapath can be found in \texttt{lib/dpif-netdev.h/c} files. 
%
To better understand the resurgence property (including its period, magnitude, etc.), and the nature (and its possible effects) of the sub-table ranking process, we refer to the source code for explanations.
We identified that, in \texttt{lib/dpif-netdev.c}, the function \texttt{static void dpcls\_sort\_subtable\_vector(struct dpcls \*cls)} sorts the sub-tables according to the number of hits each sub-table has. 
Furthermore, it is called in every second as defined by the constant called \texttt{DPCLS\_OPTIMIZATION\_INTERVAL} (in milliseconds). 

For maintaining the sub-tables, OVS uses a data structure termed as \texttt{PVECTOR}.
To iterate through this \texttt{PVECTOR}, a \texttt{MACRO} function is defined. 
According to the comments in the related source code (cf.~\cref{src:pvector}), the last element inserted (\texttt{elem2}) is seen first by the iterating function. 
\begin{lstlisting}[label=src:pvector,caption=Iteration in the tuple space,style=ovs_source]
/*    pvector_insert(&my_pvector, &elem1, 1);
*     pvector_insert(&my_pvector, &elem2, 2);
*     ...
*     PVECTOR_FOR_EACH (iter, &my_pvector) {
*         ...operate on '*iter'...
*         ...elem2 to be seen before elem1...   */
\end{lstlisting}

To better understand the impact of this behavior on the packet classification, we first have to see how a new tuple is created.
Notably, in \cref{src:create-subtable}, we see that whenever a new sub-table is created (for a flow), it is inserted into the tuple space by exactly the same function mentioned above, i.e., by the \texttt{pvector\_insert()}.

\begin{lstlisting}[label=src:create-subtable,caption=Subtable Creation,style=ovs_source]
static struct dpcls_subtable * 
dpcls_create_subtable(struct dpcls *cls, const struct netdev_flow_key *mask) {
...
/* Add the new subtable at the end of the pvector (with no hits yet) */
pvector_insert(&cls->subtables, subtable, 0); 
...}
\end{lstlisting}

We conclude the effects of the ranking process as follows.
\begin{itemize}
    \item The sub-tables are ranked according to the number of hits a particular tuple has.
    \item Sub-tables with the highest ranks are located at the end of the data structure \texttt{PVECTOR}, where the linear search process of the algorithm starts looking up the tuples.
    \item A newly created sub-table is always ranked the first, i.e., it will be added to the end of the \texttt{PVECTOR} right after it has been spawned.
\end{itemize}




\subsection{Experimental Verification}
\label{sec:mfc_ovsdpdk__improvement__experimental_verification}
To practically verify our findings, we conduct a set of experiments to make sure the time when resurgence starts and its period are constant and affected by the ranking itself.
First, we examine whether the resurgence is a one-off event by making several iterations in the same environment. 
Then, we show what the correlation between the attack rate and resurgence times is. 
Hereafter, according to the performance of \ovsdpdk we measure above (\cref{sec:evaluation__ovsdpdk}), we exclusively focus on the \texttt{SIP\_SP\_DP} attack.

\subsubsection{Justifying resurgence}
\label{sec:mfc_ovsdpdk__improvement__experimental_verification__resurgence}
First, we use the same constant attack rate of $1000$ pps as we used before, and repeat the measurements four times.
In Fig.~\ref{fig:resurgence_confirmation}, we can observe that the victim throughput always resurges ${\sim}24-26$ seconds after the attack is started, i.e., at around the $45^{\text{th}}$ second.

On the other hand, to better understand to what extent the gradually increasing number of masks affects the overall performance of \ovsdpdk, after the attack has started (i.e., after the $20^{\text{th}}$ second), each densely dotted light-grey line indicates that a subsequent batch of thousand masks has been generated.
For instance, the first dotted line (at the $21^{\text{th}}$ second) denotes that the first batch of $1000$ masks has been spawned, while the last $10^{\text{th}}$ line (at the $30^{\text{th}}$ second) marks that all possible ${\sim}9k$ masks are present in the MFC after sending the whole \texttt{SIP\_SP\_DP} packet trace (cf.~\cref{sec:evaluation__methodology_and_threat_model}).
We observe (in the magnified rectangle in Fig.~\ref{fig:resurgence_confirmation}) that, on average, the complete denial-of-service period of the attack is reached when ${\sim}7k$ masks are spawned in the MFC, i.e., the throughput becomes $0$ after the $7^{\text{th}}$ dotted light-grey line.
This finding will be crucial in developing TSE 2.0 (see details in~\cref{sec:tse_2.0}).


\subsubsection{The effect of the attack rate on resurgence}
\label{sec:mfc_ovsdpdk__improvement__experimental_verification__effect_of_attack_rate}
Next, we run the \texttt{SIP\_SP\_DP} use case with various attack rates and measure Time-To-Decay (\textit{TTD}), i.e., how much time does it take to reduce the victim throughput to zero after the attack has been launched.
Furthermore, we measure the Time-To-Resurge (\textit{TTR}), i.e., the time the resurge to the degraded service of $2.5-3.5$ Gbps happens after the attack has been started.
Last but not least, we also measure the Denial-of-Service period (\textit{DoSP}) w.r.t. the time when the attack has started, i.e., the period for which the victim throughput is entirely down. 

For \texttt{TTD}, first, consider \cref{fig:resurgence_confirmation} again, where the attack rate $1000$ pps in all cases.
To better visualize how the gradually increasing number of masks affects the overall performance of \ovsdpdk, after the attack has started (i.e., after the $20^{\text{th}}$ second), each densely dotted light-grey line indicates that a subsequent batch of thousand masks has been generated.
For instance, the first dotted line (at the $21^{\text{th}}$ second) denotes that the first batch of $1000$ masks has been spawned in one second, while the last $10^{\text{th}}$ line (at the $30^{\text{th}}$ second) marks that all possible ${\sim}9k$ masks are present in the MFC after sending the whole \texttt{SIP\_SP\_DP} packet trace (cf.~\cref{sec:evaluation__methodology_and_threat_model}).
We observe (in the magnified rectangle in Fig.~\ref{fig:resurgence_confirmation}) that, on average, the complete denial-of-service period of the attack is reached when ${\sim}6-7k$ masks are spawned in the MFC, i.e., $\texttt{TTD}=6-7$s.
On the other hand, \texttt{TTR} and \texttt{DoSP} become ${\sim}24$ and ${\sim}19$, respectively.
Intuitively, we can observe (not shown in \cref{fig:resurgence_confirmation}) that as the attack rate increases, the \texttt{TTD} decreases, i.e., 
when the attack rate is, say, $2000$, and $3000$ pps, the TTD becomes ${\sim}3.5$, and ${\sim}1.5$ seconds, while the \texttt{TTR} (and \texttt{DoSP}) become ${\sim}21.5$ (${\sim}19$), and ${\sim}19$ (${\sim}17.5$) seconds, respectively.

However, as soon as the maximum number of masks is reached (i.e., $\sim9k$), the attack becomes ``redundant''; the ranks of each sub-table corresponding to the attacker's flows become low and stable (i.e., not fluctuating heavily by constantly spawning a new mask with the highest rank). 
In contrast, the victim throughput, which is considerably higher in rate, starts to rise through the ranks, and eventually, we can observe the resurgence. 
The reason behind the victim throughput not resurging fully to its baseline is that the tuple space is still being heavily updated, i.e., the data structure still has to re-rank and relocate almost $10k$ sub-tables in every second (cf.~\cref{sec:mfc_ovsdpdk__ranking__source_code}).
Furthermore, classifying the low rate attack traffic becomes now the bottleneck; the linear search process for the attack packets still takes considerably more time, which impacts the overall packet processing performance per second.

We conclude that the performance of \ovsdpdk (w.r.t. the victim traffic) heavily depends on the following properties:
\begin{enumerate}[label=\textit{Prop(\arabic*)}, leftmargin=.7in]
    \item \label{prop:rank} the ranks of the victim's flows,
    \item \label{prop:num_of_masks} the number of masks in the MFC,
    \item \label{prop:attack_rate} the rate of the attack traffic.
\end{enumerate}

\section{TSE 2.0 against Single-Core \ovsdpdk}
\label{sec:tse_2.0}
We have seen that the ranking process in \ovsdpdk defeats the purpose of TSE 1.0 to a certain extent, i.e., only a degradation-of-service can be achieved in the long term.
Here, we extend and improve TSE 1.0 to accommodate the new enhanced mechanisms in \ovsdpdk. 

The critical aspect of the ranking process we exploit here is that as soon as a packet spawns a new sub-table, it will always be placed at the end of the tuple space, i.e., at the end of the \texttt{PVECTOR} (cf.~\cref{src:pvector}), where the linear search process starts.
Consequently, besides keeping the considerably high number of masks active, we have to create new masks continually at the same time.
However, the attack vector of TSE 1.0 is given, i.e., we can only spawn ${\sim}9k$ masks with the \texttt{SIP\_SP\_DP} trace, and the idle timeout period for each mask is $10$ seconds. 
Therefore, our idea is to let several thousands of masks expire by stopping to send their corresponding packets towards OVS for a while; then, we start to resend them again to defeat~\ref{prop:rank}.
Although, at the same time, we have to keep a significant number of other masks active in the MFC to defeat~\ref{prop:num_of_masks}.
In our proposed TSE 2.0, therefore, we fix the attack rate to as low as possible, i.e., to $1000$ pps. Therefore \ref{prop:attack_rate} has only a constant impact on the performance of \ovsdpdk.
Later, we show that if \ovsdpdk can scale up to multiple cores (~\cref{sec:tse_2.1}), \ref{prop:attack_rate} becomes increasingly important.

\subsection{Overview of TSE 2.0}
\label{sec:tse_2.0__overview}
Next, we describe how TSE 2.0 works.
The low attack rate of TSE 1.0 (i.e., $1000$ pps) enables us to generate all possible masks (${\sim}9k$) within less than the idle timeout period of $10$ seconds resulting in that no mask can expire. 
Therefore, in TSE 2.0, we choose $1000$ pps as an attack rate as well, and after $10$ seconds, i.e., after all, masks have been generated, we pause the attack to let the masks spawned in the beginning to expire.
Then, by restarting the attack and respawning the corresponding masks, they will be ranked top again, resulting in that the victim flows require again more time to be found during the lookup process.
For brevity, we term the first attacking period as \textit{attack time} ($T_{attack}$), while the pause period as \textit{sleep time} ($T_{sleep}$).
Note, $T_{sleep}$ has to be carefully chosen to last long enough to let a portion of the masks expire, but not too long as it would eventually leave only a small amount of sub-tables in the tuple space not beating \ref{prop:num_of_masks}. 

\subsection{Evaluation of TSE 2.0}
\label{sec:tse_2.0__evaluation}
Next, we experiment with several combinations of the attack and sleep times to find the optimal setting which maintains just enough number of masks in the MFC to inflate the tuple space and, at the same time, continuously (re-)generates the required number of masks rendering the sub-tables of the victim traffic ranked significantly lower. 
In particular, we evaluate the following configurations considering the most devastating traffic trace and the number of masks it can spawn and the expiration of time of each of them:
\begin{enumerate}[label=\textit{Conf(\arabic*)}, leftmargin=.7in]
    \item \label{conf:10_1} $T_{attack}=10$s, $T_{sleep}=1$s,
    \item \label{conf:10_2} $T_{attack}=10$s, $T_{sleep}=2$s,
    \item \label{conf:10_3} $T_{attack}=10$s, $T_{sleep}=3$s,
    \item \label{conf:9_2} $T_{attack}=9$s, $T_{sleep}=2$s.
\end{enumerate}

In particular, the reasons behind these configurations are the following.
First, increasing $T_{attack}$ above $10$ seconds would not facilitate the attack as the number of masks we generate is bounded by the traffic trace; all of them are generated within $10$ seconds when the attack rate is $1000$ pps.
Second, increasing $T_{sleep}$ to more than $3$ allows three thousand masks to expire, which would eventually lead to too few sub-tables present in the MFC not beating \ref{prop:num_of_masks}. 
Similarly, decreasing the attack time to only $8$ seconds has multiple consequences w.r.t. to the chosen sleep time.
In particular, if $T_{sleep}=1$ second, then restarting the attack in the $10^{\text{th}}$ second would be too early, and none of the sub-tables would be expired.
Increasing $T_{sleep}$ to $2$, however, would result in an insufficient number of masks active in the MFC at a time, i.e., the number of masks would be just around $7k$\footnote{Recall, to spawn all ${\sim}9k$ masks, we need to send all ${\sim}9.5k$ packets of the \texttt{SIP\_SP\_DP} trace that requires $9.5$ seconds in case of $1000$ pps.}, which is the bound when the attack itself reaches complete DoS.
Hence, the main question we are looking for is how many new sub-tables we have to continuously \mbox{(re-)generate} while keeping more than $7k$ sub-tables active simultaneously.

\begin{figure}[t!]
    \centering
    \subfloat[$T_{attack}=10$ and $T_{sleep}=1$ seconds.]
    {
	    \includegraphics[width=.475\textwidth]{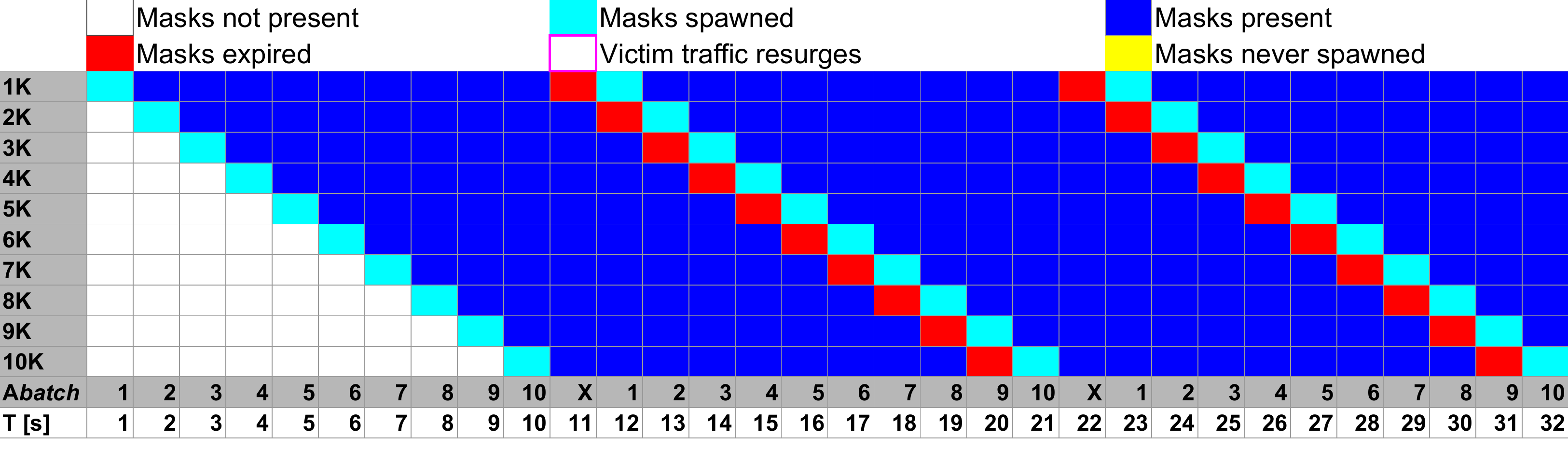}
	    \label{fig:10_1}
    }

    \subfloat[$T_{attack}=10$ and $T_{sleep}=2$ seconds.]
    {
    	\includegraphics[width=.475\textwidth]{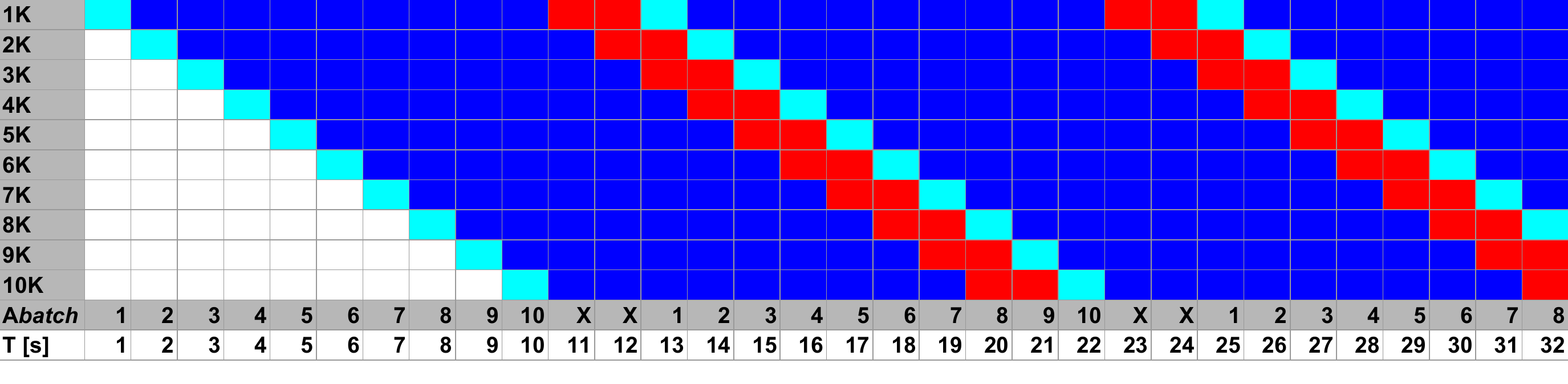}
    	\label{fig:10_2}
    }

	\subfloat[$T_{attack}=10$ and $T_{sleep}=3$ seconds.]
	{
		\includegraphics[width=.475\textwidth]{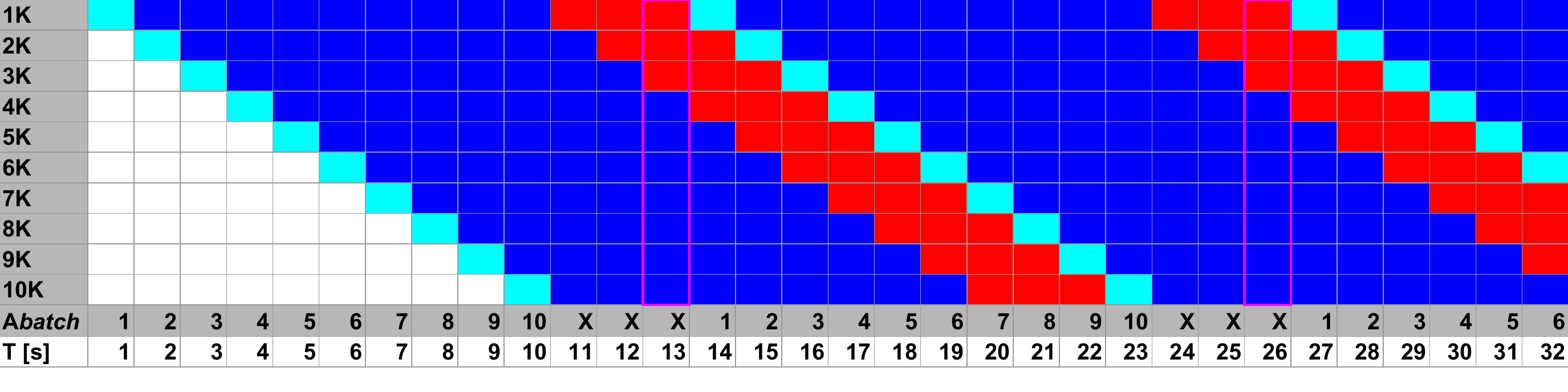}
		\label{fig:10_3}
	}

	\subfloat[$T_{attack}=9$ and $T_{sleep}=2$ seconds.]
	{
		\includegraphics[width=.475\textwidth]{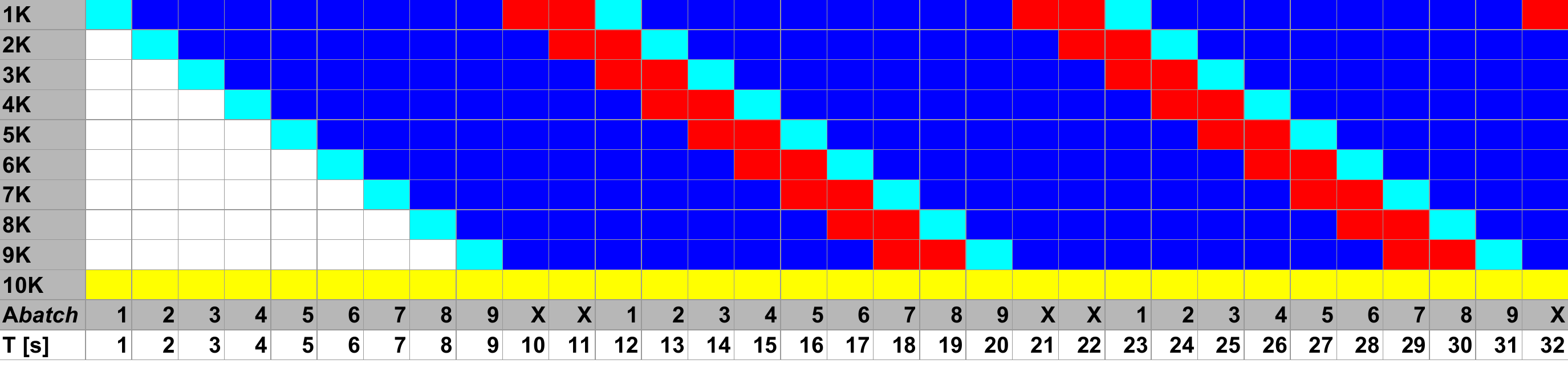}
		\label{fig:9_2}
	}
    \caption{MFC Cache maps for different configurations of TSE 2.0. Mind the common legend on the top of \cref{fig:10_1}.}
    \label{fig:mfc_cache_maps}
\end{figure}

\noindent
\textbf{MFC Cache Map. }
We display several snapshots of the MFC for each configuration as a function of time.
We refer to this representation as the \textit{MFC Cache Map} that, in particular, shows the following information.
For each second (denoted by the bottom-most row in \cref{fig:mfc_cache_maps}, i.e., by \textbf{T[s]},), we show ten cells (for brevity) to indicate the actual population of the MFC; each cell in each column represents $1000$ sub-tables, the topmost being the first \textbf{1k} while the row denoted by \textbf{10k} in the left-most column is the last \textbf{10k}.
A cell with white color means that the corresponding $1k$ masks are not present yet, while the cyan color represents that they are being generated in the given second. Note again, due to the packet trace and the attack rate, all possible sub-tables are populated after ${\sim}9.5$ seconds, and the $10^{\text{th}}$ batch of $1k$ sub-tables will not be fully populated.
A cell's color is blue if the corresponding sub-tables have been already generated, and they are still present in the MFC, i.e., they did not expire yet.
On the other hand, a cell with red color means that the corresponding $1k$ masks are expiring in that particular second.
Finally, when a cell's color is yellow (in \cref{fig:9_2} only), it means that the corresponding $1000$ sub-tables are not created at all in the given attack configuration.

The second row from the bottom ($\textbf{A}_{\textbf{\textit{batch}}}$) denotes the state of the attack.
Notably, the state $i \in [1,10]$ marks that the attack is currently sending the $i^{\text{th}}$ batch of $1000$ packets, while states $X$ show when the attack is in the \textit{sleep phase}, i.e., no packets are sent.
Last but not least, when all ten cells for a given second is highlighted by a pink bracket (in \cref{fig:10_3} only), it means that the victim traffic starts to resurge in that time slot.

The results for \ref{conf:10_1}, \ref{conf:10_2}, \ref{conf:10_3}, and \ref{conf:9_2} are depicted in Fig.~\ref{fig:10_1}, Fig.~\ref{fig:10_2}, Fig.~\ref{fig:10_3}, and Fig.~\ref{fig:9_2}, respectively.
In case of \ref{conf:10_1}, we observe that by pausing the attack for $1$ second, a batch of 1000 masks expires at every second then are re-spawned in the consecutive time interval, thereby achieving ${\sim}8k$ active masks in the MFC and a continuous \mbox{(re-)generation} of ${\sim}1k$ masks.
This approach achieves a complete DoS and never lets the victim throughput to resurge.
Since in contrast to the original TSE attack, the overall attack rate is now reduced by $1000$ pps in every $10$ seconds, the attack rate of this variant of TSE 2.0 is $596$ kbps compared to $656$ kbps\footnote{Considering the smallest possible packets (64B) plus the 20-byte overhead of the inter-frame gap and the preamble, $1000$ pps = $1000\times84\times8$ bps.}.

Next, we evaluate \ref{conf:10_2}, which lets twice the amount of more masks expire compared to \ref{conf:10_1}; however, it regenerates ${\sim}1k$ masks in each second as well after the sleep times.
The results (cf.~Fig.~\ref{fig:10_2}) shows that this approach is also able to achieve a complete DoS by not letting the victim traffic to resurge. 
Furthermore, due to longer sleep time, the average attack rate becomes even lower ($546$ kbps).

On the other hand, increasing the sleep time to $3$ seconds in \ref{conf:10_3} results in a resurgence of the victim traffic as it lets ${\sim}3k$ masks expire but it only (re-)generates ${\sim}1k$ at every second (during the attack phases) resulting in only ${\sim}6k$ active masks in the MFC at any given time. 

Finally, we discuss the results for \ref{conf:9_2} depicted in Fig.~\ref{fig:9_2}.
Recall that due to the traffic trace and $T_{attack}$, to generate all possible masks, the attack should last for around ${\sim}9.5$ seconds (cf.~\cref{sec:background__tse}), however, we observe that with  $T_{attack}=9$ seconds and $T_{sleep}=2$ seconds, we keep the required number of masks active in the MFC\footnote{Note, capturing the actual MFC population promptly is challenging due to the non-real-time update time of the user-space MFC monitoring application (i.e., \texttt{ovs-appctl}), non-deterministic optimization attempts in the MFC to produce the widest possible masks, etc.} while also letting ${\sim}2k$ masks expire and ${\sim}1k$ be regenerated keeping the ranking process busy.
Thus, \ref{conf:9_2} not only hinders the victim throughput from resurging, but it also requires the lowest attainable throughput on average ($<550$ kbps).

\section{TSE 2.1 against  Multi-Core \ovsdpdk}
\label{sec:tse_2.1}
All our improvements in \cref{sec:tse_2.0} are introduced to defeat the resurgence of the victim throughput in the case of a single-core \ovsdpdk.
In this section, we explore to what extent \ovsdpdk running on multiple cores are vulnerable to the improved TSE 2.0 attack.
In particular, as a baseline measurement, we scrutinize the impact of TSE 1.0 and TSE 2.0 on \ovsdpdk running on two cores.
In case of TSE 2.0, we choose \ref{conf:10_2} as a baseline, which is the lowest rate configuration that surely spawns all possible masks our trace \texttt{SIP\_SP\_DP} can produce (cf.~\cref{sec:tse_2.0__evaluation}).
After observing that \ovsdpdk with more resources are more robust against both variants of the TSE attack presented so far, we put more emphasis on \ref{prop:attack_rate}, i.e., we scrutinize what the required attack rate that would have a significant impact on the linear search process in the MFC is. 
However, the higher the attack rate, the faster the sub-tables are spawned.
High attack rate can quickly render the sleep times of TSE 2.0 useless as within $T_{attack}$ the sub-tables will be updated more than once, thereby restarting the timeout value of $10$ seconds (see more details in~\cref{sec:tse_2.1__preliminary_evaluation}).
Therefore, we develop TSE 2.1, where we carefully increase the attack rate and adjust packet sending sequence of the underlying \texttt{SIP\_SP\_DP} packet trace (by packet cloning) at the same time to achieve a complete DoS. 

\subsection{Preliminary Evaluation with TSE 1.0 and TSE 2.0}
\label{sec:tse_2.1__preliminary_evaluation}


We now provide an overview of the original TSE 1.0 and the improved TSE 2.0 attack against \ovsdpdk running on two cores. 
Our purpose is to evaluate the efficacy of both TSE attacks if we allow \ovsdpdk to scale.

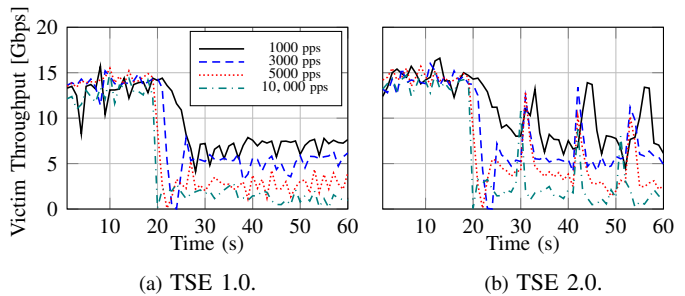
\begin{figure}[t!]
	\vspace{-1.5em}
	\subfloat[\small{TSE 1.0.}]
	{
		\hspace{-1em}
		\begin{tikzpicture}[spy using outlines={rectangle, magnification=2.4, connect spies}]
			\begin{axis}[
				width=.6\linewidth,
				height=4cm,
				xmin=1,
				ymin=0,
				ymax=20,
				xmax=60,
				xlabel={Time (s)},
				ylabel={Victim Throughput [Gbps]},
				xtick={0,10,20,30,40,50,60},
				mark=yes,
				mark size = 0pt,
				legend style={
					at={(.98,.98)},
    				anchor=north east,
					column sep=0.15cm,
					row sep=-0.15cm,
					legend columns=1,
					font=\tiny,
				},
				grid style=dashed,
				every tick label/.append style={
					font=\footnotesize
				},
				every axis y label/.style={
					at={(-0.1,0.48)},
					anchor=south,
					rotate=90,
					font=\footnotesize
				},
				every axis x label/.style={
					at={(0.5,-0.3)},
					anchor=south,
					font=\footnotesize,
				},
				ymajorgrids=true,
				xmajorgrids=true,
				grid style=solid,
				]
				\addplot [black, solid, line width=.6]   
				table[x=seconds, y={TSE_1.0_1k}, col sep=comma, mark=none]{figs/OVS_DOS_2_Core_1K_new_fixed.csv};
				\addlegendentry{$1000$ pps}
				
				\addplot [blue, densely dashed, line width=.6]  
				table[x=seconds, y={TSE_1.0_3k}, col sep=comma, mark=none]{figs/OVS_DOS_2_Core_1K_new.csv};
				\addlegendentry{$3000$ pps}
				
				\addplot [red, densely dotted, line width=.6]  
				table[x=seconds, y={TSE_1.0_5k}, col sep=comma, mark=none]{figs/OVS_DOS_2_Core_1K_new.csv};
				\addlegendentry{$5000$ pps}
				
				\addplot [teal, dashdotted, line width=.6]  
				table[x=seconds, y={TSE_1.0_10k}, col sep=comma, mark=none]{figs/OVS_DOS_2_Core_1K_new.csv};
				\addlegendentry{$10,000$ pps}
				
				
				\coordinate (spypoint) at (axis cs:23,2.5);
				\coordinate (magnifyglass) at (axis cs:10,5);
				
			\end{axis}
			
			
		\end{tikzpicture}
		\label{fig:2core_tse1.0}
		
	}
	\subfloat[\small{TSE 2.0.}]
	{
		\hspace{-1.5em}
		
		\begin{tikzpicture}
			\begin{axis}[
				width=.6\linewidth,    
				height=4cm,
				xmin=1,
				ymin=0,
				xmax=60,
				ymax=20,
				xlabel={Time (s)},
				xtick={0,10,20,30,40,50,60},
				yticklabels={},
				legend pos=north east,
				mark=yes,
				mark size = 0pt,
				legend style={
					column sep=0.15cm,
					legend columns=1,
					font=\tiny,
				},
				grid style=dashed,
				every tick label/.append style={
					font=\footnotesize
				},
				every axis y label/.style={
					at={(-0.1,0.48)},
					anchor=south,
					rotate=90,
					font=\footnotesize
				},
				every axis x label/.style={
					at={(0.5,-0.3)},
					anchor=south,
					font=\footnotesize,
				},
				ymajorgrids=true,
				xmajorgrids=true,
				grid style=solid,
				]
				\addplot [black, solid, line width=.6]   
				table[x=seconds, y=TSE_2.0_1k, col sep=comma, mark=none]{figs/OVS_DOS_2_Core_1K_new.csv};
				
				\addplot [blue, densely dashed, line width=.6]  
				table[x=seconds, y=TSE_2.0_3k, col sep=comma, mark=none]{figs/OVS_DOS_2_Core_1K_new.csv};
				
				\addplot [red, densely dotted, line width=.6]  
				table[x=seconds, y=TSE_2.0_5k, col sep=comma, mark=none]{figs/OVS_DOS_2_Core_1K_new.csv};
				
				\addplot [teal, dashdotted, line width=.6]  
				table[x=seconds, y=TSE_2.0_10k, col sep=comma, mark=none]{figs/OVS_DOS_2_Core_1K_new.csv};
			\end{axis}
		\end{tikzpicture}
		\label{fig:2core_tse2.0}
	}
	\caption{The impact of TSE 1.0 and TSE 2.0 with various attack rates against \ovsdpdk on two cores. Mind common legend.} 
	\label{fig:2core}
\end{figure}

The results are depicted in Fig.~\ref{fig:2core}, where both TSE 1.0 and TSE 2.0 are configured to send the \texttt{SIP\_SP\_DP} attack trace with a rate of $1000$ pps (black solid lines).
We observe that none of the TSE variants can degrade the victim throughput to close to $0$ Gbps.
In particular, on average, TSE 1.0 degrades the performance to ${\sim}50\%$ of the baseline. 
Similarly, TSE 2.0 reaches the same level of degradation, but, at the same time, it lets the victim throughput resurge close to its baseline throughput (${\sim}15$ Gbps) every time during the sleep phases.
This latter varying property of TSE 2.0 is a direct consequence of not being sufficiently effective with the primary attack settings, i.e., the number of masks we generate with the rate of 1000 masks per second does not impose a sufficient barrier for the linear search process when \ovsdpdk runs on two cores.
Furthermore, whenever the attack is in its sleep phase, the underlying ranking process re-rank the sub-tables more quickly (again, due to the increased resources) leading to an immediate resurgence of the victim throughput.

Consequently, to cause a complete DoS 
we need to pay more attention to \ref{prop:attack_rate}.
However, while \naively increasing the attack rate definitely imposes more impact on the performance (as more packets have to be classified per second), at the same time, we would make the resurgence of the victim throughput to happen earlier. 
In particular, consider TSE 1.0 with different attack rates in Fig.~\ref{fig:2core_tse1.0}, namely $3000$ pps (blue dashed line), $5000$ pps (red dotted line), and $10,000$ pps (teal dash-dotted line), where we can make two important observations.
First, since all possible sub-tables are generated faster, the victim traffic also starts to resurge earlier, e.g., with $10,000$ pps, the victim throughput resurges (to ${\sim}2 Gbps$) in the first second after the attack was started.
Second, the higher the attack rate, the lower the (resurged) victim throughput; however, complete DoS is not achieved even with $10,000$ pps. 

In case of TSE 2.0,  besides the similar overall impact on the victim throughput, the increased attack rate introduces further side effects (cf.~Fig.~\ref{fig:2core_tse2.0}).
Particularly, since for $3000$, $5000$, and $10,000$ pps the attainable number of masks is reached within $3$, $2$, and $1$ seconds, respectively, all sub-tables will be updated more than once within $10$ seconds. 
Thus none of them will ever expire (irrespective of the sleep periods), just like in case of TSE 1.0.
For instance, in case of $3000$ pps, all sub-tables are updated at the $9^{\text{th}}$ second within $T_{attack}=10s$, therefore, a sleep time of more than $9$ seconds would be needed to let the first batch of masks expire.
Consequently, whenever an attack phase ends, the only tasks OVS has to do in the MFC is to re-rank the sub-tables for the victim flows without needing to deal with expiring and regenerated ones.
As we observe in Fig.~\ref{fig:2core_tse2.0}, this even allows the victim throughput to rise back close to its baseline during these sleep periods.

\begin{figure*}[ht!]
	\vspace{-1.5em}
	\subfloat[\small{\ovsdpdk on 2 cores.}]
	{
		\begin{tikzpicture}
			\begin{axis}[
				width=0.35\linewidth,   
				height=4.5cm, 
				xmin=1,
				ymin=0,
				xmax=60,
				ymax=20,
				xlabel={Time (s)},
				xtick={0,10,20,30,40,50,60},
				ylabel={Victim Throughput [Gbps]},
				legend pos=north east,
				mark=yes,
				mark size = 0pt,
				legend style={
					at={(.5,.95)},
					anchor=center,
					column sep=0.15cm,
					legend columns=2,
					font=\scriptsize,
				},
				grid style=dashed,
				every tick label/.append style={
					font=\small
				},
				every axis y label/.style={
					at={(-0.1,0.48)},
					anchor=south,
					rotate=90,
					font=\small
				},
				every axis x label/.style={
					at={(0.5,-0.275)},
					anchor=south,
					font=\small,
				},
				ymajorgrids=true,
				xmajorgrids=true,
				grid style=solid,
				]
				\addplot [black, solid, line width=.6]   
				table[x=seconds, y=TSE 2.1, col sep=comma, mark=none]{figs/OVS_DOS_2_Core_1K.csv};
				\addlegendentry{3000 pps}
				
			\end{axis}
		\end{tikzpicture}
		\label{fig:TSE_2.1_2cores}
	}
	\subfloat[\small{\ovsdpdk on 3 cores.}]
	{
		\begin{tikzpicture}
			\begin{axis}[
				width=0.35\linewidth,    
				height=4.5cm, 
				xmin=1,
				ymin=0,
				xmax=60,
				ymax=20,
				xlabel={Time (s)},
				xtick={0,10,20,30,40,50,60},
				yticklabels={},
				legend pos=north east,
				mark=yes,
				mark size = 0pt,
				legend style={
					at={(.5,.95)},
					anchor=center,
					column sep=0.15cm,
					legend columns=2,
					font=\scriptsize,
				},
				grid style=dashed,
				every tick label/.append style={
					font=\small
				},
				every axis y label/.style={
					at={(-0.1,0.48)},
					anchor=south,
					rotate=90,
					font=\footnotesize
				},
				every axis x label/.style={
					at={(0.5,-0.275)},
					anchor=south,
					font=\footnotesize,
				},
				ymajorgrids=true,
				xmajorgrids=true,
				grid style=solid,
				]
				\addplot [black, solid, line width=.6]   
				table[x=seconds, y={4000 pps}, col sep=comma, mark=none]{figs/DPDK_tse2.1_3core.csv};
				\addlegendentry{$4000$ pps}
				
				\addplot [blue, densely dashed, line width=.6]  
				table[x=seconds, y={5000 pps}, col sep=comma, mark=none]{figs/DPDK_tse2.1_3core.csv};
				\addlegendentry{$5000$ pps}
				
				\addplot [red, densely dotted, line width=.6]  
				table[x=seconds, y={6000 pps}, col sep=comma, mark=none]{figs/DPDK_tse2.1_3core.csv};
				\addlegendentry{$6000$ pps}
				
				
			\end{axis}
		\end{tikzpicture}
		\label{fig:TSE_2.1_3cores}
	}
	\subfloat[\small{\ovsdpdk on 4 cores.}]
	{
		\begin{tikzpicture}
			\begin{axis}[
				width=0.35\linewidth,    
				height=4.5cm, 
				xmin=1,
				ymin=0,
				xmax=60,
				ymax=20,
				xlabel={Time (s)},
				xtick={0,10,20,30,40,50,60},
				yticklabels={},
				legend pos=north east,
				mark=yes,
				mark size = 0pt,
				legend style={
					at={(.5,.95)},
					anchor=center,
					column sep=0.15cm,
					legend columns=2,
					font=\scriptsize,
				},
				grid style=dashed,
				every tick label/.append style={
					font=\small
				},
				every axis y label/.style={
					at={(-0.1,0.48)},
					anchor=south,
					rotate=90,
					font=\footnotesize
				},
				every axis x label/.style={
					at={(0.5,-0.275)},
					anchor=south,
					font=\footnotesize,
				},
				ymajorgrids=true,
				xmajorgrids=true,
				grid style=solid,
				]
				\addplot [black, solid, line width=.6]   
				table[x=seconds, y={6000 pps}, col sep=comma, mark=none]{figs/DPDK_tse2.1_4core.csv};
				\addlegendentry{$6000$ pps}
				
				\addplot [blue, densely dashed, line width=.6]  
				table[x=seconds, y={8000 pps}, col sep=comma, mark=none]{figs/DPDK_tse2.1_4core.csv};
				\addlegendentry{$8000$ pps}
				
				\addplot [red, densely dotted, line width=.6]  
				table[x=seconds, y={10000 pps}, col sep=comma, mark=none]{figs/DPDK_tse2.1_4core.csv};
				\addlegendentry{$10000$ pps}
				
				\addplot [teal, dashdotted, line width=.6]  
				table[x=seconds, y={12000 pps}, col sep=comma, mark=none]{figs/DPDK_tse2.1_4core.csv};
				\addlegendentry{$12000$ pps}
				
			\end{axis}
		\end{tikzpicture}
		\label{fig:TSE_2.1_4cores}
	}
	
	\caption{The impact of TSE 2.1 (with \ref{conf:10_2}) with different attack rates against \ovsdpdk running on multiple cores.}
	\label{fig:TSE_2.1}
\end{figure*}
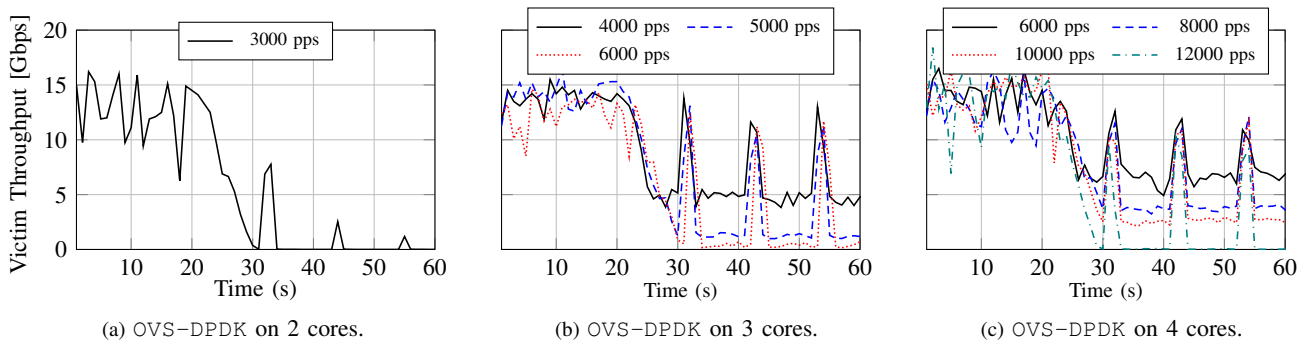

\subsection{Overview of TSE 2.1}
\label{sec:tse_2.1__overview}
In TSE 2.1, we carefully increase the attack rate and, simultaneously, adjust the sending sequence of the underlying \texttt{SIP\_SP\_DP} packet trace (by packet cloning) to achieve a complete DoS against \ovsdpdk running on multiple cores.

\algrenewcommand\algorithmicindent{1.0em}%
\begin{algorithm}[t!]
	\small
	\caption{TSE 2.1, \textit{Input:} packet trace $\mathcal{T}$,  clone factor $n$.} 
	\begin{algorithmic}[1]
		\For{\textbf{each} $p \in \mathcal{T}$} \COMMENT{we send each packet}
			\For{$i \gets 1$ to $n$}		\COMMENT{$n$ times, $n=\tfrac{A}{1000}$}
			\State $\textsc{send}(p)$
			\EndFor
		\EndFor
	\end{algorithmic}
	\label{alg:tse2.1}
\end{algorithm}

The main idea behind sending the packet trace differently is to clone and replay each packet $n$ times right after each other, where $n=\ceil{\tfrac{\text{attack~rate}}{1000}}$.
When a packet occurs for the first time, it spawns a new mask, while the consecutive \mbox{$n-1$} packets introduce a small \textit{delay} until the next, previously unseen packet is received.
By introducing delays in the sub-table generation, we can maintain a higher attack rate with the same sub-table generation rate of TSE 2.0.

To express this numerically, we define Mask Generation Rate (\texttt{MGR}) as the number of masks spawned within $1$ second. 
In the original setting, when the attack rate is, say, $3000$ pps, we generate $3000$ masks in $1$ second; hence, \texttt{MGR} is $3000$. 
This means that every new packet is received as well as every new mask is generated in every $\tfrac{1}{\texttt{MGR}} = \tfrac{1}{3000}s=0.333~ms$. 
In TSE 2.1, however, by cloning and replaying each packet $n=3$ times, \texttt{MGR} becomes $1000$ again making every new mask to be spawned \textit{only} in every $3\times\tfrac{1}{3000}s=1~ms$. 
By attaining the same \texttt{MGR} of TSE 2.0 but with three times more attack rate, we expect that the same attack and sleep time configuration causes a complete DoS.

\subsection{Evaluation of TSE 2.1}
\label{sec:tse_2.1__evaluation}

\subsubsection{\ovsdpdk with two cores}
\label{sec:tse_2.1__evaluation__2cores}
The results are shown in \cref{fig:TSE_2.1_2cores}, where the attack rate is $3000$ pps, $n=3$, while $T_{attack}=10$s and $T_{sleep}=2$s.
We observe that right after the attack has started, even though the $3000$ pps, to degrade the performance to $0$, i.e., to spawn all masks, ${\sim}10$ seconds are needed (just like in the case of $1000$ pps).
Then, besides some negligible spikes during the sleep periods, a complete DoS is achieved. 
Note, \naively reducing the sleep time is insufficient to evade these small pikes as it would not allow enough masks to expire and would not pose the required overhead of sub-table management, e.g., ranking (cf.~\cref{sec:tse_2.0__evaluation}).

\subsubsection{\ovsdpdk with three cores}
\label{sec:tse_2.1__evaluation__3cores}
Next, we analyze to what extent we need to further increase the attack rate to beat \ovsdpdk with three cores.
In Fig.~\ref{fig:TSE_2.1_3cores}, the impact of TSE 2.1 is shown with different attack rates, namely $4000$ (black solid line), $5000$ (blue dashed line), and $6000$ (red dotted line) pps with $n$ configured accordingly in each case. 
As expected, if more resources are available for OVS to scale up to, the higher the attack rate is needed to significantly degrade its performance.
In particular, compared to the case of \ovsdpdk running on two cores, the attack rate has to be doubled (to $6000$ pps) to degrade the victim throughput close to $0$.
On the other hand, we also observe a behavior similar to \cref{sec:tse_2.1__evaluation__2cores}; there are spikes during sleep times.
However, the spikes are much more significant, which can be attributed to the increased number of cores.
Note, again, eliminating the sleep phases would not evade the spikes.

Since during the attack phase (with $6000$ pps) a complete DoS is reached, and the victim throughput never resurges back to its baseline for a longer constant period than the sleep time (i.e., for a period the attacker is not in control of), we consider this attack successful.

\subsubsection{\ovsdpdk with four cores}
\label{sec:tse_2.1__evaluation__4cores}
Finally, we evaluate the performance of \ovsdpdk running on four cores.
Similarly to~\cref{sec:tse_2.1__evaluation__3cores}, we observe in Fig.~\ref{fig:TSE_2.1_4cores} that the increased number of cores requires the same increase in the attack rate. 
In particular, for bringing down \ovsdpdk on four cores, we have to increase the attack rate up to four times more than the attack rate sufficient against \ovsdpdk on two cores, i.e., up to $12,000$ pps.
Note again, during the sleep times, the victim throughput resurges, but it drops down to $0$ right after the attack period restarts.

Overall, the results can be summarized as follows.
First, in contrast to TSE 2.0, \textit{TSE 2.1 is effective against \ovsdpdk running on multiple cores} by increasing the attack rate but carefully tailoring the packet sending sequence to keep the Mask Generation Rate (\texttt{MGR}) at a constant pace of $1000$. 
Second, the more cores \ovsdpdk can scale up to, the higher the attack rate needed for a complete DoS. 
In particular, in our system (cf.~Table~\ref{tab:hw_sw_details}), we observe a linear correlation between the number of cores \ovsdpdk uses and the required attack rate, i.e., for core numbers 2, 3, and 4, the required attack rate is $3000$, $6000$, and $12,000$ pps, respectively.
Accordingly, we confirmed that to beat \ovsdpdk running on more than four cores, e.g., on five cores, the required attack rate would be $24,000$ pps, which also doubles as the number of cores assigned to \ovsdpdk increases (not shown shown for brevity).
However, as mentioned in~\cref{sec:tse_2.1__preliminary_evaluation}, as the attack rate goes beyond $15,000$ pps, we do not consider an attack to be low-rate anymore, which is the fundamental goal of the Tuple Space Explosion attack~\cite{ovs_dos}.


\section{Conclusion}
Fast packet classification is an essential part of almost every network function; however, software implementations still face performance challenges and \textit{have} corner cases.
Recently, we have presented the Tuple Space Explosion (TSE) attack~\cite{ovs_dos}, which exploiting an algorithmic deficiency in the Tuple Space Search (TSS) packet classification algorithm can degrade the performance of Open vSwitch (OVS) to \mbox{${\sim}1\%$} of its baseline with less than \mbox{$1$ Mbps} attack traffic.

In this paper, we carried out a comprehensive study to analyze the feasibility of TSE.
Notably, we showed that against the most recent OVS datapath implemented by the Linux kernel developers, TSE is indeed effective (as shown in~\cite{ovs_dos}). 
However, in contrast to our experiments in~\cite{ovs_dos}, when we use the kernel datapath module developed by the OVS developers, depending on the available resources, 
the additional caching layer can help to mitigate the TSE attack.

We also analyzed a more enhanced version of OVS, where the datapath is implemented in the user space via DPDK libraries. 
We experimentally demonstrated that due to the enhanced ranking process in the tuple space \ovsdpdk has introduced, the TSE attack has turned to be less effective. 
Therefore, we proposed TSE 2.0, which 
by letting some tuples expire and re-spawning them by carefully switching the original TSE attack on and off keeps 
the ranking process busy, 
eventually causing a complete denial-of-service (DoS) for the rest of users of the same software switch.

Furthermore, we proposed TSE 2.1 against \ovsdpdk running on multiple cores, wherein we slightly increase the attack rate of TSE 2.0 but, at the same time, we carefully adjust the packet sending sequence. 
We experimentally showed that TSE 2.1 can still mount a low-rate DoS attack as long as \ovsdpdk is running on less than five cores.

\noindent
\textbf{Acknowledgement. }
This research is supported by the National Research Foundation, Prime Minister’s Office, Singapore under its Corporate Laboratory@University Scheme, National University of Singapore, and Singapore Telecommunications Ltd.

%
%



%
\bibliographystyle{plain}
\bibliography{main.bib}

\begin{thebibliography}{10}

\bibitem{Netflow_short}
C.~{Balantrapu} et~al.
\newblock {A Novel Approach to NetFlow Monitoring in Data Center Networks}.
\newblock In {\em COMSNETS}, 2014.

\bibitem{docker_vs_lxc}
T.~Banerjee.
\newblock {Understanding the Key Differences Between LXC and Docker}.
\newblock \url{https://bit.ly/3mHQ8dG}, Aug 2014 [Accessed: Aug 2020].

\bibitem{ovs-dpdk-ranking}
B.~Bodireddy et~al.
\newblock {OVS-DPDK Datapath Classifier}.
\newblock Intel Blogpost, \url{https://intel.ly/3kCbIi8}, October 2016
  [Accessed: May 2020].

\bibitem{set_pruning_tries}
Y.~{Chang} et~al.
\newblock {Set Pruning Segment Trees for Packet Classification}.
\newblock In {\em 2011 IEEE AINA}, 2011.

\bibitem{ovs_dos}
L.~Csikor et~al.
\newblock {Tuple Space Explosion: A Denial-of-Service Attack against a Software
  Packet Classifier}.
\newblock ACM CoNEXT ’19, 2019.

\bibitem{tuplemerge}
J.~Daly et~al.
\newblock Tuplemerge: Fast software packet processing for online packet
  classification.
\newblock {\em IEEE/ACM ToN}, 2019.

\bibitem{vpp}
{FD.io}.
\newblock {VPP - Vector Packet Processing}.
\newblock \url{https://docs.fd.io/vpp/19.01/index.html}.

\bibitem{ovs-dpdk}
R.~Giller.
\newblock {Open vSwitch* with DPDK Overview}.
\newblock Intel Blogpost, \url{https://intel.ly/35TyvBO}, Sept 2016 [Accessed:
  May 2020].

\bibitem{hierarchical_cuts}
P.~{Gupta} et~al.
\newblock {Classifying Packets with Hierarchical Intelligent Cuttings}.
\newblock {\em IEEE Micro}, 2000.

\bibitem{ovs-dpdk-perf}
T.~F. Herbert.
\newblock {FD.io/VPP OVS DPDK: An Comparison of Fd.io and OVS/DPDK (sic!)}.
\newblock DPDK Summit 2016, \url{https://bit.ly/2G0AMA7}, 2016.

\bibitem{stateful_nat1}
{Intel}.
\newblock Network function virtualization.
\newblock \url{https://intel.ly/3iOCyD2}, Aug 2014 [Accessed: Jul 2020].

\bibitem{ovs-dpdk-performance-yahoo}
Intel.
\newblock {How Yahoo! JAPAN Used Open vSwitch* with DPDK to Accelerate L7
  Performance in Large-Scale Deployment Case Study}.
\newblock Blogpost, \url{https://intel.ly/2ZZYKml}, Jun 2017 [Accessed: Aug
  2020].

\bibitem{dpdk-performance}
Intel.
\newblock {DPDK Intel NIC Performance ReportRelease 20.05}.
\newblock Performance Report, \url{https://bit.ly/33NSgYL}, Jun 2020 [Accessed:
  Aug 2020].

\bibitem{dpdk}
L.~Jill.
\newblock {Data Plane Development Kit (DPDK) Further Accelerates Packet
  Processing Workloads}.
\newblock DPDK announcement, \url{https://bit.ly/3cinqLF}, Jun 2018 [Accessed:
  Jul 2020].

\bibitem{ovs_in_linux}
S.~M. Kerner.
\newblock {Open vSwitch (OVS) Becomes a Linux Foundation Collaborative
  Project}, Aug 2016 [Accessed: Jun 2020].

\bibitem{hypervisor_switch_best_practice1}
E.~Koblentz.
\newblock {Server Virtualization Best Practices and Tips on What Not to Do}.
\newblock Blogpost, \url{https://tek.io/2RN8tYS}, Jul 2018 [Accessed: Aug
  2020].

\bibitem{Kogan:2014:SAX:2619239.2626294}
K.~Kogan et~al.
\newblock {SAX-PAC: Scalable And eXpressive Packet Classification}.
\newblock In {\em ACM SIGCOMM}, 2014.

\bibitem{low_rate_dos_tifs3}
J.~Luo et~al.
\newblock {On a Mathematical Model for Low-Rate Shrew DDoS}.
\newblock {\em IEEE TIFS}, 2014.

\bibitem{low_rate_dos_tifs2}
G.~{Macia-Fernandez} et~al.
\newblock {Mathematical Model for Low-Rate DoS Attacks Against Application
  Servers}.
\newblock {\em {IEEE TIFS}}, 2009.

\bibitem{ksoftirqd}
man7.org.
\newblock {SysAdmin Series – What Does ksoftirqd Do?}
\newblock \url{https://bit.ly/2ZWDVbt}, [Accessed: Jun 2020].

\bibitem{emc}
B.~O'Mahony.
\newblock {The Open vSwitch* Exact-Match Cache}.
\newblock Intel Blogpost, \url{https://intel.ly/3iR9wma}, Aug 2017 [Accessed:
  May 2020].

\bibitem{openflow_spec}
{ONF}.
\newblock {Open Flow Switch Specification: Version 1.5.1}.
\newblock Whitepaper, \url{https://bit.ly/3hZIuYV}, 2015 [Accessed: Jun 2020].

\bibitem{ovssourcecode}
openvswitch.
\newblock {ovs}.
\newblock Github, \url{https://bit.ly/2EkL2Tz}, [Accessed: Jul 2020].

\bibitem{ovs-ludicruous}
J.~Pettit.
\newblock {Accelerating Open vSwitch to “Ludicrous Speed}.
\newblock {Blog post,\url{https://bit.ly/33S6dFn}}, 2014.

\bibitem{ovs}
B.~Pfaff et~al.
\newblock {The Design and Implementation of Open vSwitch}.
\newblock In {\em USENIX NSDI}, 2015.

\bibitem{harp}
F.~Pong et~al.
\newblock {Hashing Round-down Prefixes for Rapid Packet Classification}.
\newblock In {\em USENIX ATC}, 2009.

\bibitem{hypercuts}
Y.~Qi et~al.
\newblock {Packet Classification Algorithms: From Theory to Practice}.
\newblock IEEE INFOCOM, 2009.

\bibitem{hyperswitch}
K.~K. Ram et~al.
\newblock {Hyper-Switch: A Scalable Software Virtual Switching Architecture}.
\newblock In {\em USENIX ATC}, 2013.

\bibitem{packet_classification_computational_approach}
A.~Rashelbach et~al.
\newblock {A Computational Approach to Packet Classification}.
\newblock {\em ACM SIGCOMM}, 2020.

\bibitem{ovs-dpdk-ranking-patch}
J.~Scheurich.
\newblock {dpif-netdev: dpcls per in\_port with sorted subtables}.
\newblock commit history, \url{https://bit.ly/33Lr7G2}, July 2016 [Accessed:
  May 2020].

\bibitem{TSS}
V.~Srinivasan et~al.
\newblock {Packet Classification Using Tuple Space Search}.
\newblock {\em ACM SIGCOMM CCR}, page 135–146, August 1999.

\bibitem{kubernetes-ovn}
{The Open vSwitch project}.
\newblock {Kubernetes Integration for OVN}.
\newblock \url{https://bit.ly/3kzzNGc}.

\bibitem{openstack-ovn}
{The OpenStack project}.
\newblock {OpenStack Neutron Integration with OVN}.
\newblock \url{https://bit.ly/2ZSdiEn}.

\bibitem{less_throughput_with_emc}
A.~Theurer.
\newblock {Testing the Performance Impact of the Exact Match Cache}.
\newblock In {\em OVS Fall Conference}, 2018.

\bibitem{linux_kernel_networking_perf}
A.~Toonk.
\newblock {Linux Kernel and Measuring Network Throughput}.
\newblock Blog, \url{https://bit.ly/3hNZ5id}, Mar 2020 [Accessed: Jul 2020].

\bibitem{efficuts}
B.~Vamanan et~al.
\newblock {EffiCuts: Optimizing Packet Classification for Memory and
  Throughput}.
\newblock {\em ACM SIGCOMM CCR}, 2010.

\bibitem{low_rate_dos_tifs1}
Y.~{Xiang} et~al.
\newblock {Low-Rate DDoS Attacks Detection and Traceback by Using New
  Information Metrics}.
\newblock {\em {IEEE TIFS}}, 2011.

\bibitem{low-rate-dos_percentage}
W.~{Zhijun} et~al.
\newblock {Low-Rate DoS Attacks, Detection, Defense, and Challenges: A Survey}.
\newblock {\em IEEE Access}, 2020.

\end{thebibliography}

\begin{IEEEbiography}
[{\includegraphics[width=.9in,height=1.25in,clip,keepaspectratio]{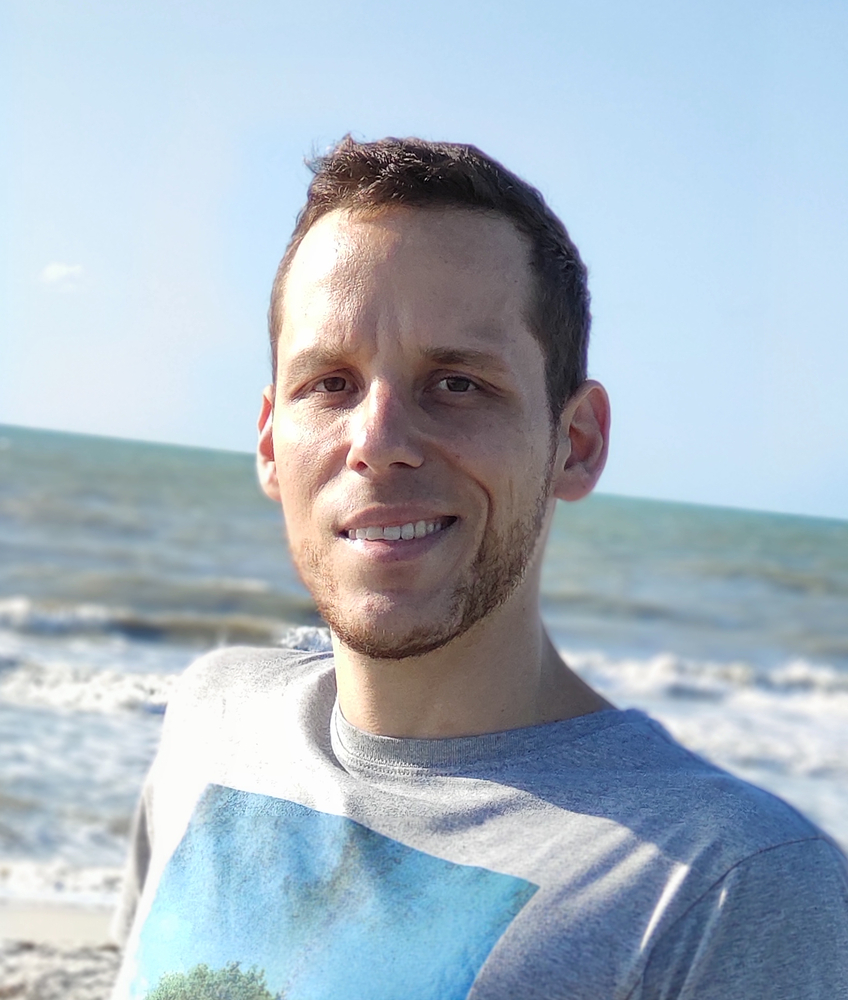}}]{Levente
Csikor} is a Senior Research Fellow at the NUS-Singtel Cyber Security R\&D Corp. Lab at the National University of Singapore.
Before joining NUS, Levente was a Research Associate at INTRIG, University of Campinas (Brazil), at the School of Computing Science, University of Glasgow (UK), as well as a visiting researcher at the Eotvos Lorand University (Hungary).
He received his M.Sc. and Ph.D. degree from Budapest University of Technology and Economics (Hungary), in 2010 and 2015, respectively.
His interests include data plane performance of different software-based network functions, network programmability, machine learning, denial-of-service attacks and privacy.
\end{IEEEbiography}

\begin{IEEEbiography}
[{\includegraphics[width=.9in,height=1.25in,clip,keepaspectratio]{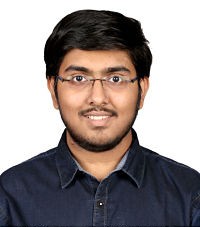}}]{Vipul Ujawane} is a final year undergraduate student at the Indian Institute of Technology(IIT), Kharagpur graduating in 2021. He was an intern at the NUS-Singtel Cyber Security R\&D Corp. Lab at the National University of Singapore (during this work). 
His interests involve offensive security, web application security, network security and privacy.
\end{IEEEbiography}

\begin{IEEEbiography}
[{\includegraphics[width=.9in,height=1.25in,clip,keepaspectratio]{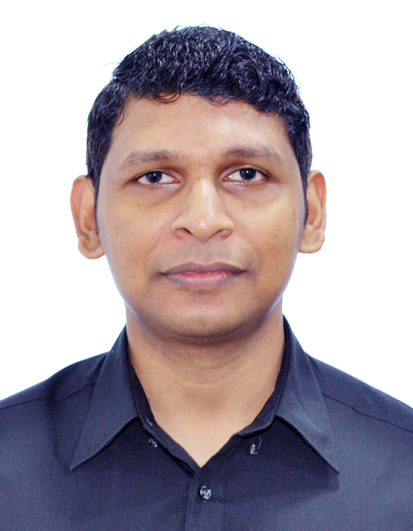}}]{Dr. Dinil Mon Divakaran} is a security researcher at Trustwave (a Singtel company). He is also an Adjunct Assistant Professor at the School of Computing in NUS (National University of Singapore). He leads and manages multiple R\&D projects at the NUS-Singtel Cyber Security R\&D Corp. Lab based in NUS. Prior to this, he was the Deputy Head of the Network Security department at the A*STAR Institute for Infocomm Research ($\text{I}^2$R). His research experience cuts across both industry and academia. He previously held faculty position at the Indian Institute of Technology (IIT) Mandi, where he joined as Assistant Professor after his PhD. He carried out his doctoral studies in ENS Lyon (France), within the joint lab of INRIA and Bell Labs. He holds a Master degree in Computer Science and Engineering from IIT Madras, India. His research interests revolve around the broad areas of network and system security, security analytics (application of stochastic models, machine learning and deep learning in security domain), protocol analysis and QoS provisioning (using programmable data planes).
\end{IEEEbiography}

\end{document}